\documentstyle[12pt,epsfig]{article}

\def\gsim{\mathrel{\rlap {\raise.5ex\hbox{$ > $}}
{\lower.5ex\hbox{$\sim$}}}}
\def\lsim{\mathrel{\rlap {\raise.5ex\hbox{$ < $}}
{\lower.5ex\hbox{$\sim$}}}}

\newcommand{\be}{\begin{equation}}
\newcommand{\ee}{\end{equation}}
\newcommand{\bea}{\begin{eqnarray}}

\newcommand{\eea}{\end{eqnarray}}

\def\gappeq{\mathrel{\rlap {\raise.5ex\hbox{$>$}}
{\lower.5ex\hbox{$\sim$}}}}

\def\lappeq{\mathrel{\rlap{\raise.5ex\hbox{$<$}}
{\lower.5ex\hbox{$\sim$}}}}

\begin{document}

\begin{titlepage}
\vspace{0.1in}

\begin{center}
{\Large {\bf
 On the  complexifications \\
of the Euclidean $R^n$ spaces
and the n-dimensional generalization of Pithagore theorem}\\

 \vspace{0.2in}

\Large{\it  Guennady Volkov}
   %\footnote{On leave from PNPI, Russia}
}

\begin{flushright}
PNPI--- St-Peterbourg\\
CERN--- Geneve\\
\end{flushright}

\end{center}

\vspace{0.2in}

\date{20.6.2010}
\vspace{0.2in}

\begin{abstract}

\vspace{0.2in}

We will discuss the following results

\begin{itemize}
\item{${C }_n$ complexification of ${ R}^n $ spaces}\\
\item{${ C}_n$ structure and the  invariant surfaces}\\
\item{${ C}_n$ holomorphicity  and harmonicity}\\
\item{ The link between ${ C}_n$ holomorphicity and the origin of spin $1/n$}\\
\item{New geometry and N-ary algebras/symmetries}\\

\end{itemize}

\end{abstract}

\end{titlepage}
\tableofcontents

\newpage
\section{Introduction. A little about History}
\begin{itemize}
\item{1.Euclid geometry  and Pithagorean theorem \cite{Euclid},\cite{Maor}};\\
\item{2. Complex numbers, Euler,s formula, complexification of
${ R}^2$, $U(1)=S^1$ \cite{Euler}}\\
\item{3.Hamilton quaternions,  octonions and geometry of unit quaternions
and octonions, the $SU(2)=S^3$ and $G(2)$ groups \cite{Hamilton, Baez}}\\
\item{4.Lie algebras and Cartan-Killing classification of  Lie algebra}\\
\item{5.Geometry of symmetric spaces and its application in physics}\\
\item{6.Complex numbers in ${ R}^3$ space. Appel sphere and
ternary generalization of trigonometric functions. \cite{Appell},
\cite{Humbert, Devisme}}\\
\item{7.The $q^n=1$ generalisation of the complex numbers in ${ R}^n$
space. \cite{FTY,FTY2},\cite{Kern}}\\
\item{8. Ternary quaternions and $TU(3)$ ternary algebra.\cite{DV}}\\
\item{9. Complex analysis in ${ R}^3$  \cite{LRV}}\\
\item{10. Calabi-Yau spaces and its algebraic classification. \cite{Berger},
\cite{AENV1, AENV2}}\\
\item{11.The reflexive number  algebra  and Berger graphs.
Its link to the n-ary Lie algebras. \cite{V}}\\
\item{12. The Standard Model problems and new n-ary algebras/symmetries.
 Searches for a new geometrical objects through the theory of new numbers. }\\
\end{itemize}

The modern progress in physics of elementary particles is based on the
discovery of the Standard Model defined by internal $SU(3) \times SU(2) \times U(1)$
gauge symmetry and external Poincar/'e symmetry. From the point of view of
the vacuum structure the SM rests on the old level, and the Higgs mechanism
 of the breaking  the $SU(2) \times U(1)$ vacuum to the $U(1)^{em}$ vacuum
does not give any geometrical picture
of the primordial vacuum.
  As the Standard Model comprises three
generations of quarks and leptons, the $SU(3) \times SU(2) \times U(1)$ symmetry cannot
fix many parameters (about 25) and cannot explain a lot of physical problems.
 The big number of the parameters inside the SM and
our non-understanding of many phenomena like families, Yukawa
interactions, fermion mass spectrum, confinement, the nature of
neutrino and its mass origin give us a proposal that the symmetry
what we saw inside the SM is only a projection of more fundamental
bigger  symmetry based on the ternary extension of the binary
Cartan-Lie symmtries. There is also an analogy with the Dark
matter problem and following this analogy we   proposed an
existence of some new  ternary  symmetries  in SM.

To understand the ambient geometry of our world with some extra
infinite dimensions one can  suggest that  our visible world
(universe) is just a subspace of a space which `` invisible'' part
one can call by bulk. The visibility of such bulk  is determined
by our understanding of the SM and our possibilities to predict
what could happened  beyond its. To find an explanation of the
small mass of neutrinos in the sea-saw mechanism it was suggested
that in this bulk could exist apart from gravitation fields some
sterile particles, like heavy right-handed neutrinos which could
interact with  light left-handed neutrinos. The Majorana neutrino
can travel in the bulk?! For this we should introduce a new space
time-symmetry which generalizes the usual D-Lorentz symmetry.

The existence of the  Majorana fermion matter  in
nature can give the further development in the understanding of
the Lorentz symmetry and  matter-antimatter symmetry,
the geometrical origin of the  gauge
symmetries of the Standard Model,   3-quark-lepton family problems,
dark matter and dark energy problems in Cosmology.
For example, embedding the Majorana  neutrino into the higher
dimensional space-time we need to find a generalization of
relativistic Dirac-Majorana equation which should not  contradict
to low energy experiments in which the properties of neutrino are
known very well!  There  could be the  different ways of embedding
the  large extra-dimensions cycles according some new symmetries,
what can give us  new  phenomena in neutrino physics, such as a
possile new SO(1,1) boost at high energies of neutrino..
The embedding the  new symmetries (ternary,...)open the window
into the extra-dimensional world with $D>3+1$, gives us
renormalizable theories in the space-time with
Dim=5,6,....similarly as Poincar/'e symmetry with internal gauge
symmetries gave the renormalizability of quantum field theories in
D=4.
The $D$-dimensional binary
Lorentz groups cannot allow to go into the $D>4$ world, {\it i.e.
} to build the renormalizable theories for the large space-time
geometry with dimension $D>4$..
It seems very plausible that using such ternary
symmetries will appear a real possibility to overcome the problems
with quantization of  a membrane theory and what could be a
further progress  beyond the string/{SS} theories. Also these new
ternary algebras could be related  with some new {SUSY}
approaches. Getting the renormalizable quantum field theories in
$D > 4$ space-time we could  find the point-like limits of  the
string and membrane theories for some new  dimensions $D >  4$.

Many interesting and important attempts have
been done to solve the problems in extensions of the Standard Model in terms of Cartan-Lie
algebras, for example: left-right extension, horizontal symmetries, $SU(5)$, $SO(10)$,
$E(6)$, $E(8)$,  Grand Unified theories, SUSY and SUGRA models.

At last in the
superstring/$D$-branes approaches it was suggested a way to construct Theory of Everything.
The theory of  superstrings is also based on the binary Lie groups,
in particular  on the
D-dimensional Lorentz group,
and therefore the description of the Standard Model in the superstrings
approach did not bring us to  success. In our opinion, one of the main problems
with the superstrings approaches  is
the inadequate exrternal symmetry at
the  string scale, $M_{\rm str} >> M_{\rm SM}$,
the D-Lorentz symmetry must
be generalized. This problem exists  also  for  GUTs.

So, all modern  theories based on the binary Lie algebras
 have a common property since the
algebras/symmetries are related with some invariant quadratic forms.

\section{From classification of Calabi-Yau spaces to the Berger graphs
and N-ary algebras}

We already know  that the superstring {GUT} s  did not  bring  us
an expected success for explanation or understanding as mentioned
above many problems of the {SM}. The main progress in superstrings
(strings) was related with understanding that we should go to the
extra dimensional geometry with $D>4$. Also the superstrings
turned us again to  study the geometrical approach, which has
brought  in XIX century the  big progress in physics. This
geometrical objects, Calabi-Yau spaces with $SU(3)$ holonomy,
appeared in the process of the compactification of the heterotic
$E(8) \times E(8)$ 10-dimensional superstring on $M_4 \otimes K_6$
space  or study the duality between 5 superstring/M/F theories.
Mathematics  \cite{Berger} discovered such objects using the
holonomy principle. To get  $K_6=CY_3$, the main constraint on the
low energy physics was to conserve a very important property of
the internal symmetry, i.e., to build a grand unified theory with
$N=1$ supersymmetry. It has been got the very important result
that the  infinite series of the compact complex $CY_n$ spaces
with $SU(n)$ holonomy can be described by algebraic way of the
reflexive numbers (projective weight vectors)
. This series   starts from the torus with complex
dimension $d=1$ and $K3$ spaces with complex dimension $d=2$, with
$SU(1)$ and $SU(2)$ holonomy groups, respectively. We would like
to stress that consideration of the extended string theories leads
us to a new geometrical objects, with more interesting properties
than the well-known symmetric homogeneous spaces using in the
SCM. For example,  the $K3=CY_2$-singularities are responsible
for producing  Cartan-Lie ADE-series matter using in the {SM}.
The singularities of $CY_n$ spaces with $n \geq 3$ should be
responsible with producing of new algebras and symmetries beyond
Cartan-Lie and which can help us to solve the questions of the SM
and {SCM} \cite{ LSVV,  DV}. This geometrical
direction is related with Felix Klein,s old ideas  in his Erlangen
program which promotes  the very closed link between geometrical
objects and symmetries.

As we already said that there are some ways to construct  ternary algebras and symmetries.
One of them  is linked to the theory of numbers.The fundamental property of the simple
 KCLA classification  is  the Abelian Cartan subalgebra and
the sircumstance that for each step generator of an algebra you can build the $su(2)$-subalgebra.
 For example, it is well known that binary complex numbers
of module $1$ are related to Abelian $U(1)=S^1$ group. The  imaginary  quaternion units are related
to the ${su(2)}$ algebra and the unit quaternions are related to the $SU(2)=S^3$ group.
 And at last octonions
are related to the $G(2)$ group.
So, our way is to  consider ternary algebras and groups based on the ternary generalization
of binary numbers
(real, complex, quaternions, octonions) (see \cite{ FTY,FTY2,  DV}).

The second geometrical way is very closed related to the  symmetries of some geometrical objects.
For example, it is well known Cartan-Lie symmetries are closely connected with homogeneous
symmetrical spaces.
Due to  the superstring approch  physiciens have got a big interet to  the
Calabi-Yau geometry. It was shown that the spaces of dimension $n=2$, $K_3$-spacesm are
closely related to the Cartan-Lie algebras. Then it was proposed that such spaces of
$n=3, 4, ...$
could be related to the new $n$-ary algebras and symmetries.  We plan to study this question
for $n=3$ case through the Berger graphs, which can be found in $CY_3$ reflexive
Newton polyhedra. We determine the Berger graphs
based on the AENV-algebraic classification of $CY_n$ spaces. Actually, the Berger graphs are
directly determined by reflexive projective weight vectors, which determine the $CY$-spaces.
The Calabi-Yau spaces with $SU(n)$ holonomy
can be studied by the algebraic way through  the integer lattice
where  one can construct  the Newton reflexive polyhedra or
the Berger graphs. Our conjecture is that the Berger graphs can be directly
related with the $n$-ary algebras. To find such algebras we
study the n-ary generalization of the well-known binary norm
division algebras, ${R}$, ${ C}$, ${ H}$,
${O}$, which helped to discover the most important "minimal"  binary simple Lie groups,
$U(1)$, $SU(2)$ and  $G(2)$. As the most important example,
 we consider
the case $n=3$, which gives  the ternary generalization of quaternions (octonions), $3^n$, $n=2,3$,
respectively.
The ternary generalization of quaternions  is directly related to the new ternary algebra (group)
which are related to the natural extensions of the  binary $su(3)$ algebra ($SU(3)$ group).

Our interest in ternary algebras and symmetries started
from the study of the geometry  based  on the  holonomy principle,
discovered by Berger \cite{Berger}. The $CY_n$ spaces
with $SU(n)$ holonomy
have a special interest for us. Our  conjecture \cite{ V, LSVV}
is that $CY_3$ ($CY_n$) spaces are related to the ternary ($n$-ary)
symmetries,
which are natural generalization
of the binary  Cartan--Killing--Lie symmetries.

The holonomy group H is one of the
main characteristic of an affine connection on a manifold M.
The definition of holonomy group is directly connected with parallel
transport along the piece-smooth path  joining  two points $x \in M$
and $y\in M$.
For a connected  n-dimensional manifold M with Riemannian metric g and
Levi-Civita connection  the parralel transport along
using the connection  defines the isometry between the scalar products
on the tangent spaces
$T_xM$ and $T_yM$ at the points x and y.
So for any
point $x \in M$ one can represent the set of all linear automorphisms
of the associated tangent spaces $T_xM$ which are induced by parallel
translation  along x-based loop.

If a connection is locally symmetric then
its holonomy group equals to the local isotropy subgroup of the isometry group G.
Hence, the holonomy group classification
of these connections is equivalent to the classification of symmetric spaces
which was done completely long ago.
The full list of symmetric spaces is given by the theory of  Lie groups
through the homogeneous spaces $M=G/H$, where G is a connected  group Lie
acting transitively on M and H is a closed connected Lie subgroup of G,
what determines the holonomy group of M.
Symmetric spaces have a transitive group of isometries. The known examples
of symmetric spaces are  ${R}^n$, spheres $S^n$, ${ { {CP}}}^n$ etc.
There is a very interesting fact that Riemannian spaces (M,g) is locally
symmetric if and only if it  has constant curvature $\nabla R=0$.

If we consider  irreducible (compact, simply-connected)
Riemannian manifolds one can find there classical manifolds, the symmetric
spaces, determined by following form $G/H$, where $G$ is a compact Lie
group and H is the holonomy group itself. These spaces are completely
classified and their geometry is well-known.
But there exists  non-symmetric irreducible Riemannian manifolds
with the following list of holonomy groups H of M.

Firstly, in 1955, Berger presented  the classification of irreducibly acting
matrix Lie groups occured as the  holonomy of a torsion free affine connection.

The set of homogeneous polynomials of degree $d$ in the complex projective
space ${{\bf CP}^n}$ defined
by the vector $\vec{k_{n+1}}$ with ${d=k_1+...k_{n+1}}$ defines a
convex  polyhedron, whose intersection with the integer
lattice corresponds to the exponents of the monomials of the  equation.
Batyrev found the properties  of such polyhedra like reflexivity
which directly links these polyhedra to the Calabi-Yau equations.
Therefore, instead of studying the complex hypersurfaces
directly, firstly, one can study the geometrical properties of such
polyhedra..

One of the main results in the Universal Calabi-Yau Algebra (UCYA)  is
that the reflexive weight vectors (RWVs) $\vec{k_n}$ of dimension $n$ can
obtained directly from lower-dimensional RWVs $\vec{k_{1}}, \ldots,
\vec{k_{n-r+1}}$ by algebraic constructions of arity
$r$~\cite{AENV1}. One of the important consequences of UCYA one can see
the lattice structure connected to the Berger  graphs.
In K3 case it was shown that the Newton reflexive polyhedra are constructed by pair of plane Berger graphs coinciding to the
Dynkin diagrams of CLA algebra. In $CY_3$ the four  dimensional reflexive polyhedra are constructed from triple of Berger
graphs which by our opinion could be related to the new algebra, which can be the ternary generalizations of
binary CLAs:
\begin{eqnarray}
\vec{k_1}=(0,...,1)[1],     \qquad  & \rightarrow & \qquad    A_r^{(1)}(K3),
\qquad    TA_r^{(1)}(CY_3), ...\nonumber\\
\vec{k_2}=(0,...,1,1)[2],   \qquad  & \rightarrow & \qquad    D_r^{(1)}(K3),
\qquad    TD_r^{(1)}(CY_3),...\nonumber\\
\vec{k_3}=(0,...,1,1,1)[3], \qquad  & \rightarrow & \qquad    E_6^{(1)}(K3),
\qquad    TE_6^{(1)}(CY_3),...\nonumber\\
\vec{k_3}=(0,...,1,1,2)[4], \qquad  & \rightarrow & \qquad    E_7^{(1)}(K3),
 \qquad    TE_7^{(1)}(CY_3),...\nonumber\\
\vec{k_3}=(0,...,1,2,3)[6]  \qquad  & \rightarrow & \qquad    E_8^{(1)}(K3),
\qquad    TE_8^{(1)}(CY_3),...\nonumber\\
\label{fivevectors}
\end{eqnarray}

So, the other important success of UCYA is that it is naturally connected
to the
invariant topological numbers, and therefore it gives correctly all
 the double-, triple-, and etc. intersections, and, correspondingly,
all graphs,
which are connected with affine algebras.

It was shown  in the toric-geometry approach
how the Dynkin diagrams of affine Cartan-Lie algebras appear in reflexive
K3 polyhedra~\cite{Bat}. Moreover, it was found in~\cite{AENV1}, using
examples of the lattice structure of reflexive polyhedra for $CY_n$:
$n \geq 2$ with elliptic fibres that there is an interesting
correspondence  between the five basic RWVs
(\ref{fivevectors}) and Dynkin diagrams for the five ADE types of Lie
algebras: A, D and E$_{6,7,8}$..
For example, these RWVs are constituents
of composite RWVs for K3 spaces, and the corresponding K3 polyhedra can be
directly constructed out of certain Dynkin diagrams, as illustrated in
. In each case, a pair of extended RWVs have an
intersection which is a reflexive plane polyhedron, and one vector from
each pair gives the left or right part of the three-dimensional reflexive
polyhedron, as discussed in detail in~\cite{AENV1}.

 One can illustrate this correspondence on the example of RWVs,
 $\vec{k_3}=(k_1,k_2,k_3)[d_{\vec k}]=(111)[3],(112)[4],(123)[6]$,
for which we show how to build the
 $ E_6^{(1)}$, $E_7^{(1)}$, $E_8^{(1)}$ Dynkin diagrams, respecrtively.
Let take the vector $\vec{k_3}=(111)[3]$. To construct the Dynkin diagram
one  should start from one common node, $V^{0}$, which will give start
to n=3 (= dimension of the  vector) line-segments. To get the number of the
points-nodes $p$ on each line one should divide
$d_{\vec k}$ on $k_i$, $i=1,2,3$, so $p_i=d_{\vec k}/k_i$
( here we considere the cases when all  divisions are integers).
One should take into account, that all lines have one common node $V^{0}$.
The numbers of the points equal to $ n  \cdot(d_{\vec k}/k_i-1)+1$.
Thus, one can check, that for all these three cases there appear the
$ E_6^{(1)}$, $E_7^{(1)}$, $E_8^{(1)}$ graphs, respectively. Moreover,
one can easily see how to reproduce for all these graphs
the Coxeter labels and the Coxeter number. Firstly, one should prescribe
the Coxeter label to the comon point $V^0$. It equals
to $ max_i \{p_i\}$. So in our three cases the maximal Coxeter label,
prescribing to the common point  $V^0$,
is equal $3,4,6,$ respectively. Starting from the Coxeter label of the
node $V^0$, one can easily find the Coxeter numbers of the rest points
in each line. Note that this rule will help us in the cases of
higher dimensional $CY_d$ with $d \geq 3$, for which one can easily represent
the corresponding polyhedron and graphs without computors.

Similarly, the huge set of five-dimensional RWVs $\vec{k_5}$ in 4242
CY$_3$ chains of arity 2 can be constructed out of the five RWVs already
mentioned plus the 95 four-dimensional K3 RWVs $\vec{k_4}$.
. In this case, reflexive 4-dimensional polyhedra
are also separated into three parts: a reflexive 3-dimensional
intersection polyhedron and `left' and `right' graphs. By
construction, the corresponding CY$_3$ spaces are seen to possess K3 fibre
bundles.

We illustrate the case of one such arity-2 K3 example \cite{AENV1, AENV2}.
. In this case, a reflexive K3 polyhedron is determined by
the two RWVs ${\vec k}_1=(1)[1]$ and ${\vec k}_3=(1,2,3)[6]$. As one can
see, this K3 space has an elliptic Weierstrass fibre, and its polyhedron,
determined by the RWV ${\vec k}_4=(1,0,0,0)+(0,1,2,3)=(1,1,2,3)[7]$, can
be constructed from two diagrams, $A_6^{(1)}$ and $E_8^{(1)}$, depicted to
the left and right of the triangular Weierstrass skeleton.
The analogous arity-2 structures of all 13 eldest K3 RWVs~\cite{AENV1}.

The  extra uncompactified dimensions make quantum field
theories with Lorentz  symmetry much less comfortable, since the power
counting is worse.
A possible way out is to suppose that the propagator is more
convergent than $1/p^2$, such a behaviour can be obtained if we
consider, instead of binary  symmetry algebra, algebras with
higher order relations (\emph{That is, instead of binary
operations such as addition or product of 2 elements, we start with
composition laws that involve at least $n$ elements of the considered
algebra, $n$-ary algebras}). For instance, a  ternary symmetry could  be
related with membrane dynamics.
To solve the Standard Model problems we suggested to generalize their
external and internal binary symmetries
by addition of ternary symmetries based on the ternary algebras
\cite{V, LSVV}.
For example,  ternary symmetries seem to give very good possibilities to
overcome  the above-mentioned problems, {\it i.e.} to make the next
progress in understanding of the space-time geometry of our Universe.
We suppose that the new symmetries beyond the well-known binary Lie
algebras/superalgebras could allow us  to build the renormalizable
theories for space-time geometry with dimension $D>4$. It seems
very plausible that using such ternary symmetries will offer a
real possibility to overcome the problems of  quantization of
membranes and  could be a further progress  beyond string
theories.

 Our interest in  the new $n$-ary algebras and their classification
started from a study of infinite series of $CY_n$ spaces characterized
by holonomy groups \cite{Berger}.
More exactly, the $CY_n$ space can be defined as the quadruple
$(M,J,g,\Omega)$, where $(M,J)$ is a complex compact n-dimensional
manifold.

A $CY_{n-2}$ space can be realized as an algebraic variety ${\mathcal M}$
in a weighted projective space
${\rm CP}^{n-1} (\overrightarrow{k})$
where the weight vector reads
$\overrightarrow{k} = (k_1, \dots, k_{n})$.

The points in
CP$^{n-1}$ satisfy the property of projective invariance
$\left\{x_1, \dots, x_{n}\right\} \approx
\left\{\lambda^{k_1} x_1, \dots, \lambda^{k_{n}} x_{n}\right\}$ leading to the
constraint
$\overrightarrow{m} \cdot \overrightarrow{k} = d_k$.

The classification of $CY_n$ can be done through
the reflexivity of the weight
vectors $\overrightarrow{k}$  (reflexive numbers), which can be defined
in terms of the Newton reflexive polyhedra \cite{Bat} or Berger
graphs \cite{V}. The Newton reflexive polyhedra are determined
by the exponents of the monomials participating in the $CY_n$ equation
\cite{Bat}. The term "reflexive" is related with the mirror duality of
Calabi--Yau spaces
and the corresponding Newton polyhedra \cite{Bat}.
The Berger graphs can be constructed directly through the reflexive weight
numbers  $\vec {k}=(k_1,...,k_{n+2})[d_k]$ by the
procedure shown in  \cite{V,LSVV}.
 According to the universal algebraic
approach \cite{AENV1}
one can find  a section in the  reflexive
polyhedron and, according to the $n$-arity of this algebraic approach,
the reflexive polyhedron can be constructed from 2-, 3-,... Berger graphs.
It was conjectured that the Berger graphs might correspond
to $n$-ary Lie algebras \cite{V, LSVV}.
In these articles we tried to decode those Berger graphs by using
the
method of the "simple roots".

All modern  theories based on the binary Lie algebras
 have the common property since the
algebras/symmetries are related with some invariant quadratic
forms. Ternary algebras/symmetries should be linked also with
certain cubic invariant forms. Our interest to the new n-ary algebras and their
classification started from study of infinite series of $CY_n$
spaces characterized by holonomy groups \cite{Berger}. More
exactly,the $CY_n$ space can be defined as the quadruple
$(M,J,g,\Omega)$, where $(M,J)$ is a complex compact n-dimensional
manifold with complex structure $J$, $g$ is a kahler metrics with
$SU(n)$ holonomy group holonomy, and $\Omega_n=(n,0)$ and $\bar
\Omega_n=(0,n)$ are non-zero parallel tensors  which called by the
holomorphic volume forms.

A CY space can be realized as an algebraic variety ${\mathcal M}$
in a weighted projective space ${\rm CP}^{n-1}
(\overrightarrow{k})$  where the weight vector
reads $\overrightarrow{k} = (k_1, \dots, k_{n})$. This variety is
defined by
\begin{eqnarray}
{\mathcal M} \equiv (\left\{x_1, \dots, x_{n} \right\} \in {\rm
CP}^{n-1} (\overrightarrow{k}): {\mathcal P} (x_1, \dots, x_{n})
\equiv \sum_{\scriptsize{\overrightarrow{m}}}
c_{\scriptsize{\overrightarrow{m}}}
x^{\scriptsize{\overrightarrow{m}}} = 0 ), \label{Bat}
\end{eqnarray}
i.e., {as} the zero locus of a quasi--homogeneous polynomial of
degree $d_k = \sum_{i=1}^{n} k_i$, with the monomials being
$x^{\scriptsize{\overrightarrow{m}}} \equiv x_1^{m_1} \cdots
x_{n}^{m_{n}}$. The points in CP$^{n-1}$ satisfy the property of
projective invariance $\left\{x_1, \dots, x_{n}\right\} \approx
\left\{\lambda^{k_1} x_1, \dots, \lambda^{k_{n}} x_{n}\right\}$
leading to the constraint $\overrightarrow{m} \cdot
\overrightarrow{k} = d_k$.

For classyfying and decoding the new graphs one can  use the following rules:

\begin{enumerate}
\item{to   classify the graphs one can do according to the arity,\it i.e.} \\
 {for arity 2 here can  be two graphs, and the points on the left (right) graph
should be  on the edges  lying on one side with respect to the arity 2 intersetion}\\
 {for arity 3 there can be three graphs, which points
can be defined with respect to the arity 3 intersections
and etc.}\\
{for arity r there can be $r$ graphs}\\

\item {The graphs should correspond to extension of affine graphs of Kac-Moody algebra}\\

\item{The graphs can correspond to an universal  algebra  with some arities}\\
\end{enumerate}

The first proposal was already discussed before.
The second proposal is important because  a possible new algebra could be connected very
closely with geometry.
Loop algebra is a Lie algebra associated to a
group of mapping from manifold   to a Lie group.
 Concretely to get affine Kac-Moody it  was considered the  case where the manifold is
 the unit
circle and group  is a matrix Lie group. Here it can be a further  geometrical way
to generalize the affine Kac-Moody algebra. We will take this in mind, but
we will always suppose that the affine property of the new graphs should remain as it was
in affine Kac-Moody algebra classification.
The affine property means that the matrices corresponding to these algebras should
have the determinant equal to zero, and all principal minors of these matrices
should be positive definite. The matrices will be constructed with  almost the same
rules as the generalized Cartan matrices in affine Kac-Moody case.
 We just make one changing
on the some diagonal elements, which can take the value not only 2, but also 3 for
$CY_3$
case (4 for $CY_4$ case and etc).
The third proposal is  connected with taking in mind that a new algebra could be an
universal
algebra,
{\it i.e. } it contains apart from binary operation also  ternary,... operations.
The suggestion  of using a ternary algebra interrelates with the topological structure
of $CP^2$. This  can be used for resolution of $CY_3$ singularities.
It seems that taking into consideration
the different dimensions,  one can understand very  deeply how to extend the notion of

Lie algebras and to construct the  so called universal algebras.
 These algebras
could play the main role in  understanding of{\it non-symmetric} Calabi-Yau
geometry and can give a further progress in the understanding of
high energy physics in the Standard model and beyond.

Our  plan is following, at first we study the graphs  connected with
five reflexive weight vectors, $(1), (11), (111,(112),(123)$
and then, we   consider the  examples with $K3$- reflexive weight vectors.

To study the lattice structure of the graphs in reflexive polyhedra  one
should recall a little bit about Cartan matrices and
Dynkin diagrams..

Our reflexive polyhedra allow us to consider new graphs, which we will
call Berger graphs, and
for corresponding  Berger matrices we suggest the folowing rules:
\begin{eqnarray}
{ B}_{ii}&=&2\qquad  or\qquad  3, \nonumber\\
{ B}_{ij}& \leq& 0,\nonumber\\
{ B}_{ij}=0 &\mapsto & {B}_{ji}=0, \nonumber\\
{ B}_{ij} &\in& {Z} ,\nonumber\\
Det { B} &=&0,\nonumber\\
Det { B}_{\{(i)\}} &>& 0.\nonumber\\
\end{eqnarray}
We call the last two restriction  the {\it affine condition}.
In these  new rules  comparing with the generalized affine Cartan matrices
 we relaxed the restriction on the diagonal element
${B}_{ii}$, {\it i.e. } to satisfy the affine conditions
we allow also to be
\begin{eqnarray}
{B}_{ii}=3\, for \, CY_3, \qquad
{B}_{ii}=4\, for \, CY_4,\qquad and \,\, etc.
\end{eqnarray}
 Apart from these rules we will check the coincidence of the
graph's labels, which
we indicate on all figures with analog of Coxeter labels,
what one can get from
getting eigenvalues of the Berger matrix.

An interesting  subclass of the reflexive numbers
is the so--called ``simply--laced'' numbers (Egyptian numbers).
A simply--laced number $\overrightarrow{k}= (k_1, \cdots, k_n)$
with degree $d = \sum_{i=1}^{n} k_i$ is defined such that
\begin{eqnarray}
\frac{d}{k_i} &\in& { Z}^+ \, \, {\rm and} \, \, d > k_i.
\end{eqnarray}
For these numbers there is a simple way of constructing the corresponding
affine Berger graphs together with their Coxeter labels\cite{V,LSVV}. The
Cartan and Berger matrices of these graphs are symmetric. In the well known
Cartan case they correspond to the $ADE$ series of simply--laced algebras.
In dimensions $n = 1, 2, 3$ the Egyptian numbers are
$(1),(1,1),(1,1,1),(1,1,2),(1,2,3)$. For $n=4$ among all 95
reflexive numbers 14 are simply--laced Egyptian numbers (see Table).

{\small \begin{table}
\vspace{-3cm}
\hspace{2cm}
\begin{tabular}{|l|c|c|c|c|} \hline
${\vec k}_{3,4}^{\rm ext}        $ & $ {\rm Rank} $&$h            $ & ${\rm Casimir} (B_{ii})
$ & $ {\rm Determinant}                                                   $  \\
\hline \hline
$(0,1,1,1)[3]      $ & $ 6 (E_6)  $ & $12         $ & $6          $ & $3  $  \\
$(0,1,1,2)[4]      $ & $ 7 (E_7)  $ & $18         $ & $8          $ & $2  $  \\
$(0,1,2,3)[6]      $ & $ 8 (E_8)  $ & $30         $ & $12         $ & $1  $  \\
$(0,0,1,1,1)[3]    $ & $ 2_3+10+l $ & $18+3(l+1)  $ & $9          $ & $3^4$  \\
$(0,0,1,1,2)[4]    $ & $ 2_3+13+l $ & $32+4(l+1)  $ & $12         $ & $4^3$  \\
$(0,0,1,2,3)[6]    $ & $ 2_3+15l  $ & $60+6(l-1)  $ & $18         $ & $6^2$  \\ \hline
$(0,1,1,1,1)[4]    $ & $ 1_3+11   $ & $28         $ & $12         $ & $16 $  \\
$(0,2,3,3,4)[12]   $ & $ 1_3+12   $ & $90         $ & $36         $ & $ 8 $  \\
$(0,1,1,2,2)[6]    $ & $ 1_3+13   $ & $48         $ & $18         $ & $ 9 $  \\
$(0,1,1,1,3)[6]    $ & $ 1_3+15   $ & $54         $ & $18         $ & $ 12$  \\
$(0,1,1,2,4)[8]    $ & $ 1_3+17   $ & $80         $ & $24         $ & $ 8 $  \\
$(0,1,2,2,5)[10]   $ & $ 1_3+17   $ & $100        $ & $30         $ & $ 5 $  \\
$(0,1,3,4,4)[12]   $ & $ 1_3+17   $ & $120        $ & $36         $ & $ 3 $  \\
$(0,1,2,3,6)[12]   $ & $ 1_3+19   $ & $132        $ & $36         $ & $ 6 $  \\
$(0,1,4,5,10)[20]  $ & $ 1_3+26   $ & $290        $ & $60         $ & $ 2 $  \\ \hline
$(0,1,1,4,6)[12]   $ & $ 1_3+24   $ & $162        $ & $36         $ & $ 6 $  \\
$(0,1,2,6,9)[18]   $ & $ 1_3+27   $ & $270        $ & $54         $ & $ 3 $  \\
$(0,1,3,8,12)[24]  $ & $ 1_3+32   $ & $420        $ & $72         $ & $ 2 $  \\
$(0,2,3,10,15)[30] $ & $ 1_3+25   $ & $420        $ & $90         $ & $ 4 $  \\
$(0,1,6,14,21)[42] $ & $ 1_3+49   $ & $1092       $ & $126        $ & $ 1 $  \\
\hline
\end{tabular}
\caption{Rank, Coxeter number $h$, Casimir depending on $B_{ii}$ and determinants
for the non--affine exceptional Berger graphs. The maximal Coxeter labels
coincide with the degree of the corresponding reflexive simply--laced vector.
The determinants in the last
column for the infinite series (0,0,1,1,1)[3], (0,0,1,1,2)[4] and (0,0,1,2,3)[6]
are independent from the number $l$ of internal binary $B_{ii}=2$ nodes.
The numbers $1_3$ and $2_3$ denote the number of nodes with $B_{ii}=3$.}
\label{BERDET}
\end{table}}

%\begin{figure}
 %  Requires
% \usepackage{graphicx}
 %\includegraphics[width=14cm, height=10cm]{5Figure.ps}\\
  %\caption{5 Berger graphs -ternary extensions of ADE Dynkin
  % diagrams}\label{5}
%end%{figure}

Let compare the binary affine Dynkin diagrams for ${\cal E}_6$ and affine
 Berger graph defined by reflexive vector $(0,1,1,1,1)$.

\begin{table}
%\scriptsize
\centering \caption{\it The simple roots of the of CLA ${\cal E}_6$ and
 Berger  algebra
defined by reflexive number $k=(0,1,1,1,1)[4]$.} \label{tabl23}
\vspace{.05in}
\begin{tabular}{||c|c||} \hline
$                $&$                    $ \\
$                $&$                    $ \\
$                $&$                    $ \\
$                $&$                    $ \\
$  {\cal E}_6    $&$ \qquad\qquad
\begin{picture}(120,20)
\thicklines
\put(56,28){\circle{10}}
\put(56,56){\circle{10}}
\put(56,5){\line(0,1){18}}
\put(56,33){\line(0,1){18}}
\multiput(5,0)(28,0){4}{\line(1,0){18}}
\multiput(0,0)(28,0){5}{\circle{10}}
\put(0,-15){\makebox(0.4,0.6){$\alpha_1$}}
\put(30,-15){\makebox(0.4,0.6){$\alpha_2$}}
\put(58,-15){\makebox(0.4,0.6){$\alpha_3$}}
\put(88,-15){\makebox(0.4,0.6){$\alpha_4$}}
\put(116,-15){\makebox(0.4,0.6){$\alpha_5$}}
\put(72,26){\makebox(0.4,0.6){$\alpha_6$}}
\put(72,56){\makebox(0.4,0.6){$-\alpha_0$}}
\end{picture} \qquad \qquad                 $ \\
$                $&$                        $ \\
$                $&$                        $ \\ \hline \hline
$                $&$                        $ \\
$                $&$                        $ \\
$                $&$                        $ \\
$                $&$                        $ \\
$                $&$                        $ \\
$                $&$                        $ \\
$  {\cal BER}    $&$ \qquad
\begin{picture}(120,20)
\thicklines
\put(84,28) {\circle{10}}\put(84,5)  {\line(0,1){18}}
\put(84,56) {\circle{10}}\put(84,33) {\line(0,1){18}}
\put(84,84) {\circle{10}}\put(84,61){\line(0,1){18}}
\put(84,-28){\circle{10}}\put(84,-5) {\line(0,-1){18}}
\put(84,-56){\circle{10}}\put(84,-33){\line(0,-1){18}}
\put(84,-84){\circle{10}}\put(84,-61){\line(0,-1){18}}
\multiput(5,0)(28,0){6}{\line(1,0){18}}
\multiput(0,0)(28,0){7}{\circle{10}}
\put(0,-15){\makebox(0.4,0.6){$\alpha_1$}}
\put(28,-15){\makebox(0.4,0.6){$\alpha_2$}}
\put(56,-15){\makebox(0.4,0.6){$\alpha_3$}}
\put(72,-10){\makebox(0.4,0.6){$\alpha_4$}}
\put(112,-15){\makebox(0.4,0.6){$\alpha_5$}}
\put(140,-15){\makebox(0.4,0.6){$\alpha_6$}}
\put(168,-15){\makebox(0.4,0.6){$\alpha_7$}}
\put(102,-28){\makebox(0.4,0.6){$\alpha_8 $}}
\put(102,-56){\makebox(0.4,0.6){$\alpha_9 $}}
\put(102,-84){\makebox(0.4,0.6){$\alpha_{10}$}}
\put(102,28) {\makebox(0.4,0.6){$\alpha_{11}$}}
\put(102,56) {\makebox(0.4,0.6){$\alpha_{12}$}}
\put(102,84) {\makebox(0.4,0.6){$-\alpha_{0}$}}
\put(0,15){\makebox(0.4,0.6){$1$}}
\put(28,15){\makebox(0.4,0.6){$2$}}
\put(56,15){\makebox(0.4,0.6){$3$}}
\put(72,10){\makebox(0.4,0.6){$4$}}
\put(112,15){\makebox(0.4,0.6){$3$}}
\put(140,15){\makebox(0.4,0.6){$2$}}
\put(168,15){\makebox(0.4,0.6){$1$}}
\put(72,-28){\makebox(0.4,0.6){$3$}}
\put(72,-56){\makebox(0.4,0.6){$2$}}
\put(72,-84){\makebox(0.4,0.6){$1$}}
\put(72,28) {\makebox(0.4,0.6){$3$}}
\put(72,56) {\makebox(0.4,0.6){$2$}}
\put(72,84) {\makebox(0.4,0.6){$1$}}
\end{picture} \qquad \qquad                    $ \\
$                $&$                           $  \\
$                $&$                           $ \\
$                $&$                           $ \\
$                $&$                           $ \\
$                $&$                           $ \\
$                $&$                           $ \\
$                $&$                           $ \\\hline \hline
\end{tabular}
\end{table}
\normalsize

\begin{eqnarray}
\alpha_1&=&e_1-e_2 \nonumber\\
\alpha_2&=&e_2-e_3 \nonumber\\
\alpha_3&=&e_3-e_4 \nonumber\\
\alpha_4&=&e_4-e_5-e_9\nonumber\\
\alpha_5&=&e_5-e_6 \nonumber\\
\alpha_6&=&e_6-e_7 \nonumber\\
\alpha_7&=&e_7-e_8\nonumber\\
\alpha_8&=&e_9-e_{10} \nonumber\\
\alpha_9&=&-\frac{1}{2}(e_9-e_{10}+e_1+e_2+e_3+e_4+e_{11}-e_{12}) \nonumber\\
\alpha_{10}&=&e_{11}-e_{12} \nonumber\\
\alpha_{11}&=&e_9+e_{10} \nonumber\\
\alpha_{12}&=&-\frac{1}{2}(e_9+e_{10}-e_5-e_6-e_7-e_8+e_{11}+e_{12})  \nonumber\\
\alpha_{13}&=&e_{11}+e_{12}=- \alpha_0\nonumber\\
\end{eqnarray}

where
\begin{eqnarray}
&&4\alpha_4+3(\alpha_3+\alpha_5+\alpha_8+ \alpha_{11})+
2(\alpha_2+\alpha_6+\alpha_9+ \alpha_{12})\nonumber\\
&&+(\alpha_1+\alpha_7+\alpha_{10}+\alpha_0)=0\nonumber\\
\end{eqnarray}

\begin{eqnarray}
  \left (
 \begin{array}{ccc|c|ccc|ccc|ccc}
  2  &-1  & 0  & 0 & 0 & 0 & 0 & 0 & 0 & 0 & 0 & 0 & 0\\
 -1  & 2  &-1  & 0 & 0 & 0 & 0 & 0 & 0 & 0 & 0 & 0 & 0\\
  0  &-1  & 2  &-1 & 0 & 0 & 0 & 0 & 0 & 0 & 0 & 0 & 0\\ \hline
  0  & 0  &-1  & 3 &-1 & 0 & 0 &-1 & 0 & 0 &-1 & 0 & 0\\ \hline
  0  & 0  & 0  &-1 & 2 &-1 & 0 & 0 & 0 & 0 & 0 & 0 & 0 \\
  0  & 0  & 0  & 0 &-1 & 2 &-1 & 0 & 0 & 0 & 0 & 0 & 0 \\
  0  & 0  & 0  & 0 & 0 &-1 & 2 & 0 & 0 & 0 & 0 & 0 & 0 \\ \hline
  0  & 0  & 0  &-1 & 0 & 0 & 0 & 2 &-1 & 0 & 0 & 0 & 0 \\
  0  & 0  & 0  & 0 & 0 & 0 & 0 &-1 & 2 &-1 & 0 & 0 & 0 \\
  0  & 0  & 0  & 0 & 0 & 0 & 0 & 0 &-1 & 2 & 0 & 0 & 0 \\ \hline
  0  & 0  & 0  &-1 & 0 & 0 & 0 & 0 & 0 & 0 & 2 &-1 & 0 \\
  0  & 0  & 0  & 0 & 0 & 0 & 0 & 0 & 0 & 0 &-1 & 2 & 0 \\
  0  & 0  & 0  & 0 & 0 & 0 & 0 & 0 & 0 & 0 & 0 &-1 & 2 \\ \hline
 \end{array}
 \right )
\nonumber
\end{eqnarray}
 Note, that the determinant is equal  $Det=4^2$. In general case for $CY_d$, $d+2=n$, which
corresponds to the RWV ${\vec k}_n=(1,\ldots, 1) [n] $, the determinant
of the corresponding non-affine matrices is equal $n^{n-2}$ ( $n\geq 3$).

\normalsize

\section{${ C}_N$-division numbers and N-ary algebras}

We want to find an example of ternary non-Abelian algebra and
 to understand the mechanism of appearing
in Cartan matrix $B_{ii}=3$. For this we will go to ther study of ternary division algebras.
Historically the discovery of Killing-Cartan-Lie algebras was
closely related to the four norm division ${ C}_2$
algebras,  ${ R}$,${ C}$,${ H}$,${
O}$, {i.e. \it} real numbers, complex numbers, quaternios and
octonions, respectively \cite{ Hamilton, Cayley, Kostr1,
Baez}.

An algebra A will be a vector space that is
equipped with a bilinear map $m:A \times A \rightarrow A$
called by
multiplication and a nonzero element $1 \in A$ called the unit such
that $m(1,a)=m(a,1)=a$.
A normed division algebra is an algebra A that
is also a normed vector space with $|ab|=|a||b|$.
So ${ R}$, ${ C}$ are  the  commutative associative normed algebras,
${H}$ is noncommutative associative normed algebra.
${O}$ are the  octonions- an
non-associative alternative algebra.
An algebra is alternative if $a(ab)=a^2b$ and $(ab)b=ab^2$
( $a(ba)=(ab)a$).  An alternative division algebras has unity and  inverse
element. The only alternative division binary algebras over ${ R}$
are ${ R}$, ${ C}$, ${Q}$, ${ O}$.

 An algebra $A$ will be a vector
space that is equipped with a bilinear map $f:A \times A
\rightarrow A$ called by multiplication and a nonzero element $1
\in A$ called the unit, such that $f(1,a)=f(a,1)=a$. These
algebras admit an anti-involution (or
 conjugation)
$(a^*)^*=a$ and $ (ab)^*={b}^*  {a}^*$.
A norm division algebra is an algebra $A$ that
is also a normed vector space with $N(ab)=N(a)N(b)$.
Such algebras exist only for $n=1,2,4,8$ dimensions where
the following identities can be obtained:
\begin{eqnarray}
(x_1^2+...+x_n^2)(y_1^2+...+y^2)=(z_1^2+...+z_n^2)
\end{eqnarray}
The doubling process, which is known as the Cayley-Dickson process,
forms  the sequence of divison algebras
\begin{eqnarray}
{ R} \rightarrow { C} \rightarrow {H}
 \rightarrow { O}.
\end{eqnarray}
 Note  that next algebra is not a division algebra. So $n=1$
${ R}$ and $n=2$ ${ C}$ these algebras are  the
commutative associative normed division algebras. The quaternions,
${ H}$, $n=4$ form the  non-commutative and associative
norm division algebra. The octonion algebra  $n=8, { O}$ is
an non-associative alternative algebra. If the discovery of
complex numbers took a long period about some centuries years, the
discovery of quaternions and octonions was made in  a short time,
in the middle of the ${\rm XIX}$ century by W. Hamilton
\cite{Hamilton}, and by J. Graves and  A.Cayley \cite{Cayley}. The
complex numbers, quaternions and octonions can be presented in the
general form:

\begin{eqnarray}
\hat q= x_0 e_0+x_pe_p,\qquad  \{x_0,\,x_p\} \in { R},
\end{eqnarray}
where $p=1$ and $e_1 \equiv {\bf i}$   for complex numbers ${ C}$,
$p=1,2,3$ for quaternions ${H}$, and $p=1,2,...,7$ for
${ O}$. The $e_0$ is as unit and all $e_p$ are imaginary units
with conjugation $\bar {e}_p=-e_0$.
 For quaternions we have the main  relation
\begin{eqnarray}
e_me_p=-\delta_{mp} +f_{mpl}e_l,
\end{eqnarray}
where $\delta_{mp}$ and $f_{mpl}\equiv \epsilon_{mpl}$ are
the well-known Kronecker and Levi-Cevita tensors, respectively.
For octonions the completely antisymmetric tensor
$f_{mpl}=1$ for the following seven triple associate cycles:
\begin{equation}
\{mpl\}=\{123\},\{145\},\{176\},\{246\},\{257\},\{347\},\{365\}.
\end{equation}

There are also 28 non-associate cycles.
Each triple accociate cycle corresponds to a quaternionic subalgebra.
These algebras have a very close link with geometry.
For example, the unit elements  $x^2=1, x \in {R}$,
$|\hat q|=x_0^2+x_1^2=1$ in $ { C}_1$,
$|\hat q|=x_0^2+x_1^2+x_2^2+x_3^2=1$ in ${ H}_1$,
$|\hat q|=x_0^2+x_1^2+...+x_7^2=1$ in ${ O}_1$,
define the spheres, $S^0$,  $S^1$, $S^3$,
$S^7$, respectively.

If the binary alternative division algebras ( real numbers,
complex numbers, quaternions, octonions) over the real numbers
have the dimensions           $2^n$, $n=0,1,2,3,4,...$, the
ternary  algebras have the following dimensions $3^n$,
$n=0,1,2,3,4,...$, respectively:

\begin{eqnarray}
\begin{array}{cc|cc}
 { R}    :& \, 2^0  \,=\,1
&{ R}    :& \, 3^0  \,=\,1                       \\
 {C}    :& \, 2^1  \,=\,1+1
&{ TC}   :& \, 3^1  \,=\,1+1+1                   \\
 { Q}    :& \, 2^2  \,=\,1+2+1
&{ TQ}   :& \, 3^2  \,=\,1+2+3+2+1               \\
 { O}    :& \, 2^3  \,=\,1+3+3+1
&{TO}   :& \, 3^3  \,=\,1+3+6+7+6+3+1           \\
 {S}    :& \, 2^4  \,=\,1+4+6+4+1
&{ TS}   :& \, 3^4  \,=\,1+4+10+16+19+16+10+4+1  \\
\end{array}
\end{eqnarray}

In the last line one can see the sedenions which are do not
produce division algebra. For both cases we have the unit element
$e_0$ and the $n$ basis elements:

\begin{eqnarray}
{ R}  \rightarrow { TC}  \rightarrow {TQ}
\rightarrow {TO} \rightarrow { TS}  \rightarrow
\ldots
\end{eqnarray}

The complex numbers is 2-dimensional algebra with basis $e_0$ and
$e_1 \equiv {\bf i}$,
\begin{equation}
{C}={ R} \oplus { R} e_1,
\end{equation}
where $e_0^2=e_0$, $e_1e_0=e_0e_1=e_1$ and $e_1$ is the imaginary
unit, ${\bf i}^2=-e_0$. Considering one additional basis imaginary
unit element $e_2\equiv {\bf j}$ in the Dickson-Cayley doubling
process one can get the quaternions,
\begin{equation}
{ H}={C} \oplus  { C} j.
\end{equation}

It means that quaternions can be considered as a pair of complex
numbers:
\begin{equation}
q=(a+ {\bf i}b)+j( c+ {\bf i} d),
\end{equation}
where
\begin{equation}
j( c+ {\bf i} d)= (c+ \bar {\bf i} d) j=(c - {\bf i}d ) j.
\end{equation}
so, one can see that $ {\bf i} {\bf j}=-{\bf j}{\bf i}={\bf k}$.

The quaternions
\begin{equation}
q=x_0e_0 +x_1e_1+x_2e_2+x_3e_3, \qquad q\in {H},
\end{equation}
produce over ${ R}$  a 4-dimensional norm division algebra
where appears the fourth imaginary unit $e_3=e_1e_2 \equiv {\bf
k}$.  The main multiplication rules of all these 4-th elements
are the following:
\begin{eqnarray}
&&{\bf i}^2={\bf j}^2={\bf k}^2=-1 \nonumber\\
&&{\bf ij }={\bf k} \qquad{\bf ji }=-{\bf k}, \nonumber\\
\end{eqnarray}
All other identities can be obtained from cyclic permutations of
${\bf i,j,k}$. The imaginary quaternions ${\bf i,j,k}$ produce the
$su(2)$ algebra. There is the matrix realization of quaternions
through the Pauli matrices:
\begin{eqnarray}
\sigma_0, \,\,{\bf i} \sigma_1, {\bf i} \sigma_2, {\bf i}
\sigma_3,
\end{eqnarray}

The unit quaternions $q=a  {\bf 1}+b {\bf i}+c {\bf j}+d {\bf k}
\in H_1$, $q \bar q=1$, produce the $SU(2)$ group:
\begin{eqnarray}
q \bar q=a^2+b^2+c^2+d^2=1, \,\, \{a,b,c,d\}\in S^3, \, S^3
\approx SU(2).
\end{eqnarray}

Simililarly,  continuing  the Cayley-Dickson doubling process
${ O}={ H} \otimes { H}$,
\begin{equation}
(x_1,x_2)(y_1,y_2)=(x_1y_1-{\bar y}_2 x_2,x_2 {\bar y}_1+y_2x_1),
\qquad (\bar(x,y)=(\bar x, \bar y),
\end{equation}
one can build the octonions:
\begin{eqnarray}
{ O} ={Q} \oplus { Q} {\bf l}, \nonumber\\
\end{eqnarray}
where we introduced new basis element ${\bf l} \equiv e_4$.

 As result of this process
the basis $\{1,{\bf i}, {\bf j}, {\bf k}\}$ of ${H}$ is
complemented to a basis $\{ {\bf 1}=e_0, {\bf i}=e_1, {\bf j}=e_2,
{\bf k}=e_3=e_1e_2, {\bf l}=e_4, {\bf il}=e_5=e_1e_4, {\bf
jl}=e_6=e_2e_4, {\bf kl}=e_7=e_3e_4   \}$ of ${O}$.
\begin{eqnarray}
{\it o} = x_0e_0+x_1e_1+x_2e_2+x_3e_3+x_4e_4+x_5e_5+x_6e_6+x_7e_7
\end{eqnarray}
where we can see the  following seven associative cycle triples:
\begin{eqnarray}
&&\{123:e_1e_2=e_3\}, \,\{145:e_1e_4=e_5\},\, \{176:e_1e_7=e_6\},\, \{246:e_2e_4=e_6\},  \nonumber\\
&&\{257:e_2e_5=e_7\}, \,\{347:e_3e_4=e_7\},\, \{365:e_3e_6=e_5\}.
\end{eqnarray}

In order to find new number algebras one can use the method of the
classification of the finite groups which is known in literature \cite{Lederman, John}.
On this way one can discover the  geometrical objects invariant on the new symmetries,
First of all, it will be useful to consider the abelian cyclic
groups, ${ C}_N=\{q^N=1| 1,q,q^2,...,q^{(N-1)} \}$ of order
$N>2$ {\it i.e.} $N=3,4,5,...$. Following to the the complex
numbers where the base unit imaginary element $i^2=-1$ we will
consider two cases: $q^N= \pm 1$. A representation of the group
$G$ is a homomorphism of this group into the multiplicative group
$GL_m(\Lambda)$ of nonsingular matrices over the field $\Lambda$,
where $\Lambda={ R}, { C}$ or etc. The degree of
representation is defined by the size of the ring of matrices. If
degree is equal one the representation is linear. For abelian
cyclic group ${ C}_N$ one can easily find the character
table, which is $N \times N$ square matrix whose rows correspond
to the different characteras for a particular conjugacy clas,
$q^{\alpha}$, $\alpha=0,1,.., N-1$. For cyclic groups ${
C}_N$ the $N$ irreducible representations are one dimensional (
see Table):
\begin{eqnarray}
\left (
\begin{array}{c|cccccc}
    -          & 1 & q               &     ... &  q^{\alpha}         &
     ...       & q^{N-1}             \\\hline
    \xi^{(1)}  & 1 & 1               &     ... &  1                  &
   ...         &  1                  \\
    \xi^{(2)}  & 1 & \xi^{(2)}_2     &     ... &  \xi^{(2)}_{\alpha} &
   ...         &  \xi^{(2)}_N        \\
   ...         &  ....               &     ... & ...                 &
   ...         & ...                 \\
   \xi^{(k)}   & 1 & \xi^{(k)}_2     &     ... &  \xi^{(k)}_{\alpha} &
   ...         &  \xi^{(k)}_N        \\
   ...         &  ....               &     ... & ...                 &
   ...         & ...                 \\
  \xi^{(N)}    & 1 & \xi^{(N)}_2     &     ... &  \xi^{(N)}_{\alpha} &
  ...          &  \xi^{(N)}_N        \\
\end{array}
\right)
\end{eqnarray}
where the chracters can be defined through  N-th root of unity.
For example, if  the character table for $C_N$ can be summarised
as
\begin{equation}
\xi^{\alpha}_k=\xi^{\alpha}\exp \{(2 \pi i (k-1) (\alpha -1)
)/N\}, \qquad (k, \alpha=1,1,2,...,N).
\end{equation}

Let us consider some examples.

We remind that for the cyclic group $\mathit{C}_{2}$ there are two
conjugation classes, $1$ and $i$ and two one-dimensional
irreducible representations:

\[
\begin{array}{c|cc|c}
\mathit{C}_{2} & 1 & i &  \\ \hline
R^{(1)} & 1 & 1 & z \\
R^{(2)} & 1 & -1 & \bar{z}
\end{array}
\,\,.
\]
The cyclic group $\mathit{C}_{3}$ has three conjugation classes,
$q_{0}$, $q$ and $q^{2}$, and, respectively, three one dimensional
irreducible representations, ${R}^{(i)}$, $i=1,2,3$. We write down
the table of their characters, $\xi _{l}^{(i)}$:

\begin{eqnarray}
\left (
\begin{array}{c|ccc}
    -         & 1 & q    & q^2  \\\hline
    \xi^{(1)} & 1 & 1    &  1   \\
    \xi^{2)}  & 1 & j    & j^2  \\
    \xi^{(3)} & 1 & j^2  & j    \\
\end{array}
\right)
\end{eqnarray}

for ${ C}_3$  ( $j_3=\equiv j=\exp \{2 \pi /3 \}$).

The cyclic group $\mathit{C}_{4}$ has four conjugation classes,
$q_{0}$, $q$, $q^2$ and $q^{3}$, and, respectively, four one dimensional
irreducible representations, ${R}^{(i)}$, $i=1,2,3,4$. We write down
the table of their characters, $\xi _{l}^{(i)}$:
\begin{eqnarray}
\left (
\begin{array}{c|cccc}
    -         & 1 & q    & q^2 &  q^3  \\\hline
    \xi^{(1)} & 1 & 1    &  1  &  1    \\
    \xi^{(2)} & 1 & i    & -1  & -i    \\
    \xi^{(3)} & 1 &-1    &  1  & -1   \\
    \xi^{(4)} & 1 &-i    & -1  &  i   \\
\end{array}
\right)
\end{eqnarray}

for ${C}_4$ (( $j_4=\exp \{ \pi /2\}$),

Correspondingly, the cyclic group $\mathit{C}_{6}$ has six conjugation classes,
$q_{0}$, $q$,...,$q^{5}$, and, respectively,six one dimensional
irreducible representations, ${R}^{(i)}$, $i=1,2,3,...,6$. We write down
the table of their characters, $\xi _{l}^{(i)}$:
\begin{eqnarray}
\left (
\begin{array}{c|cccccc}
    -         & 1 & q      & q^2   &  q^3   & q^4    & q^5   \\\hline
    \xi^{(1)} & 1 & 1      &  1    &  1     & 1      & 1     \\
    \xi^{(2)} & 1 & j_6    &  j_6^2&  j_6^3 & j_6^4  & j_6^5 \\
    \xi^{(3)} & 1 & j_6^2  &  j_6^4&  1     & j_6^2  & j_6^4 \\
    \xi^{(4)} & 1 & j_6^3  &  1    &  j_6^3 & 1      & j_6^3 \\
    \xi^{(5)} & 1 & j_6^4  &  j_6^2&  1     & j_6^4  & j_6^2 \\
    \xi^{(6)} & 1 & j_6^5  &  j_6^4&  j_6^3 & j_6^2  & j_6  \\
\end{array}
\right)
\end{eqnarray}

and for ${ C}_6$, ($j_6  =\exp\{\pi i /3\}$),
respectively.

For all examples one can see the orthogonality relations:
\begin{equation}
<\xi^{(k)}, \xi^{(l)}>= \delta_{kl}.
\end{equation}

 To check this for the ${ C}_6$ case   one should take into
account the next identities:

\begin{eqnarray}
&&1+j_6+j_6^2+j_6^3+j_6^4+j_6^5=0   \nonumber\\
&&j_6+j_6^3+j_6^5=0,\,\,\, j_6-j_6^2=1,  \nonumber\\
&&1+j_6^2+j_6^4=0,\, \,\,j_6^5-j_6^4=1,\nonumber\\
\end{eqnarray}
or
\begin{eqnarray}
&&j_6  =\frac{1}{2} +i \frac{\sqrt{3}}{2}, \qquad
j_6^2=\frac{-1}{2}+i \frac{\sqrt{3}}{2}, \qquad
j_6^3=-1, \nonumber\\
&&j_6^4=\frac{-1}{2} -i \frac{\sqrt{3}}{2}, \qquad
j_6^5=\frac{1}{2}-i \frac{\sqrt{3}}{2}, \qquad
j_6^6=1.\nonumber\\
\end{eqnarray}

We confined ourselves by the case $C_6$ cyclic group since we supposed to solve
the neutrino problem using the consideration of the ${ R}^6$ space.

So, the main idea is to use the cyclic groups ${ C}^n$
and new N-ary algebras/symmetries
to find the new geometrical "irreducible"' substructures in
${ R}^n$ spaces, which are not the  consequences of the simple extensions
of the known structures of Euclidean ${R}^2$ space.

For the ternary complexification of the vector space, ${R}^{3}$,
one uses its cyclic symmetry subgroup $\mathit{C}_{3}={R}_{3}$
\cite{mc1, mc2}. In the physical context the elements of the group
$\mathit{C}_{3}$ are actually spatial rotations through a
restricted set of angles, $0,2\pi /3,4\pi /3$ around, for example,
the $x_{0}$-axis. After such rotations the coordinates,
$x_{0},x_{1},x_{2}$, of the point in ${R}^{3}$ are linearly
related with the new coordinates, $x_{0}^{\prime },x_{1}^{\prime
},x_{2}^{\prime }$ which can
be realized by the $3\times 3$ matrices corresponding to the $\mathit{C}_{3}$%
-group transformations. The vector representation $D^{V}$ is
defined through the following three orthogonal matrices:
\[
R^{V}(q_{0})=O(0)=\left(
\begin{array}{ccc}
1 & 0 & 0 \\
0 & 1 & 0 \\
0 & 0 & 1
\end{array}
\right) \,,
\]

\[
R^{V}(q)=O(2\pi /3)=\left(
\begin{array}{ccc}
1 & 0 & 0 \\
0 & -1/2 & \sqrt{3}/2 \\
0 & -\sqrt{3}/2 & -1/2
\end{array}
\right) \,,
\]

\[
R^{V}(q^{2})=O(4\pi /3)=\left(
\begin{array}{ccc}
1 & 0 & 0 \\
0 & -1/2 & -\sqrt{3}/2 \\
0 & \sqrt{3}/2 & -1/2
\end{array}
\right) \,.
\]

These matrices realize the group representation due to the relations $%
R^{V}(q^{2})=(R^{V}(q))^{2}$ and $(R^{V}(q))^{3}=R^{V}(q_{0})$.
The representation is faithful because the kernel of its
homomorphism consists only of identity: $KerR=q_{0}\in
\mathit{C}_{3}$.

Let us introduce the matrix

\begin{eqnarray}
\hat x = x_i \cdot R^V(q_i)= \left (
\begin{array}{ccc}
x_0+x_1+x_2 & 0 & 0 \\
0 & x_0-1/2(x_1+x_2) & -\sqrt{3}/2(x_1-x_2) \\
0 & \sqrt{3}/2(x_1-x_2 ) & x_0-1/2(x_1+x_2)
\end{array}
\right ).
\end{eqnarray}

The determinant of this matrix is

\begin{eqnarray}
Det (\hat x)=x_0^3+x_1^3+x_2^3-3x_0x_1x_2\,.
\end{eqnarray}
where $R^{(1)}$ is the trivial representation, whereby each
elements is mapped onto unit, \textit{i.e.} for $R^{(1)}$ the
kernel is the whole group, $\mathit{C}_{3}$. For $R^{(2)}$ and
$R^{(3)}$ the kernels can be identified with unit element, which
means that they are faithful representations, isomorphic to
$\mathit{C}_{3}$.

Based on the character table one can obtain
\[
\xi ^{V}=(\xi ^{V}(q_{0}),\xi ^{V}(q),\xi ^{V}(q^{2}))=(3,0,0)\,,
\]
which demonstrates how the vector representation $R^{V}$
decomposes in the irreducible representations $R^{(i)}$:
\[
\xi ^{V}=\xi ^{(1)}+\xi ^{(2)}+\xi ^{(3)}
\]
or
\[
R^{V}=R^{(1)}\oplus R^{(2)}\oplus R^{(3)}.
\]
The combinations of coordinates on which $R^{V}$ acts irreducible
are given below

\[
\left(
\begin{array}{c}
z \\
\tilde{z} \\
\tilde{\tilde{z}}
\end{array}
\right) =\left(
\begin{array}{ccc}
1 & 1 & 1 \\
1 & j & j^{2} \\
1 & j^{2} & j
\end{array}
\right) \left(
\begin{array}{c}
x_{0} \\
x_{1}q \\
x_{2}q^{2}
\end{array}
\right) \,.
\]

\newpage
\section{Three dimensional theorem Pithagor and ternary  complexification of
${R}^3$}

To go further
here we must interprete  some results from the ideas  of the article \cite{FTY, FTY2,Kern, LRV}.
We would like to build the new numbers based on the ${C}_3$ finite  discrete group.
For this let consider two basic elements, $q_0$, $q_1$ with the following constraints:
\begin{eqnarray}
q_1 \cdot q_0=q_0 \cdot q_1=q_1,\,\,\,\, q_1^3=q_0,\nonumber\\
\end{eqnarray}
In this case one can introduce  a new element $q_2 =q_1^2=q_1^{(-1)}$,
{\it i.e.} $q_2q_1=q_1q_2=q_0$.

From these three elements one can build a new
field ${TC}$:
\begin{eqnarray}
{ TC } = { R} \oplus { R} q_1  \oplus { R} q_1^2.
\end{eqnarray}

with the new numbers

\begin{eqnarray}
z=x_0 q_0 + x_1 q_1 + x_2 q_2, \qquad x_i \in { R}, \qquad i=0,1,2,
\end{eqnarray}

which are the ternary generalization  of the complex numbers.

Let define the operation of the conjugation:
\begin{eqnarray}
\tilde q_1=j q_1, \qquad {\tilde {\tilde q}}_1 =j^2 q_1,
\end{eqnarray}
where $j=\exp {(2 {\bf i} \pi)/3}$.
Since $q_2=q_1^2$ one can easily get
\begin{eqnarray}
{\tilde q}_2=j^2 q_2, \qquad {\tilde {\tilde q}}_2 =j q_2.
\end{eqnarray}
One can apply these two conjugation operations, respectively:
\begin{eqnarray}
&&       {\tilde z} = x_0 q_0 + x_1 j   q_1 + x_2 j^2q_2, \nonumber\\
&&{\tilde  {\tilde z}}= x_0 q_0 + x_1 j^2 q_1 + x_2 j  q_2. \nonumber\\
\end{eqnarray}
Now one can introduce the cubic invariant form:
\begin{eqnarray}
<z>^3=z {\tilde z} {\tilde {\tilde z}}=x_0^3+x_1^3+x_2^3-3x_0x_1x_3.
\end{eqnarray}

One can also easily check the following identity
\begin{eqnarray}
<z_1z_2>^3=<z_1>^3<z_2>^3,
\end{eqnarray}
which indicate about a group properties of these new ${
TC}$ numbers. We suggest that this new Abelian group can be
related with a ternary group?!

According to table of characterts one can   define two operations of the conjugations:
\begin{eqnarray}
\bar q_1=j q_1, \qquad {\bar {\bar q}}_1 =j^2 q_1,
\end{eqnarray}
where $j=\exp {(2 {\bf i} \pi)/3}$.
Since $q_2=q_1^2$ we can easily obtain
\begin{eqnarray}
{\bar q}_2=j^2 q_2, \qquad {\bar {\bar q}}_2 =j q_2.
\end{eqnarray}
These two conjugation operations can thus be applied, respectively:
\begin{eqnarray}
&&       {\bar {\hat z}} = x_0 q_0 + x_1 j   q_1 + x_2 j^2q_2, \nonumber\\
&&{\bar  {\bar {\hat z}}}= x_0 q_0 + x_1 j^2 q_1 + x_2 j  q_2. \nonumber\\
\end{eqnarray}
We now introduce the cubic form:
\begin{eqnarray}
\langle{\hat z}\rangle={\hat z} {\bar {\hat z}} {\bar {\bar {\hat z}}}=x_0^3+x_1^3+x_2^3-3x_0x_1x_3,
\end{eqnarray}

The generators $q$ and $q^2$ can be represented in the matrix form:
\begin{eqnarray}
 q=
\left (
\begin{array}{ccc}
    0  &     1   &   0      \\
    0  &     0   &   j       \\
    j^2  &     0   &   0        \\
\end{array}
\right), \qquad
 q^2=
\left (
\begin{array}{ccc}
    0  &     0   &   j     \\
    1  &     0   &   0       \\
    0  &     j^2   &   0        \\
\end{array}
\right)
\end{eqnarray}
where one can introduce the ternary transposition operations:
$1\rightarrow 2 \rightarrow 3 \rightarrow 1$ and $3 \rightarrow 2 \rightarrow 1 \rightarrow 3$ .
We now introduce the cubic form:
\begin{eqnarray}
\langle{\hat z}\rangle={\hat z} {\bar {\hat z}} {\bar {\bar {\hat
z}}}=x_0^3+x_1^3+x_2^3-3x_0x_1x_3,
\end{eqnarray}

And also easily check the following relation:
\begin{eqnarray}
\langle{\hat z}_1{\hat z}_2\rangle=\langle{\hat
z}_1\rangle\langle{\hat z}_2\rangle,
\end{eqnarray}
which indicates  the group properties of the    ${ TC}$
numbers. More exactly, the unit ${ TC}$ numbers produce the
Abelian ternary group. According to the ternary analogue of the
Euler formula, the following ternary complex  functions
\cite{FTY,FTY2, Kern} can be constructed:

\begin{eqnarray}
\Psi =\exp{(q_1 \phi_1 + q_2\phi_2) },\qquad \psi_1=\exp{(q_1
\phi_1) }, \qquad \psi_2=\exp{(q_2 \phi_2) },
\end{eqnarray}
where $\phi_i$ are the  group  parameters. For  the functions
$\psi_i$, $i=0,1,2$, {\it i.e.} we have the following analogue of
Euler, formula:
\begin{eqnarray}
\Psi&=&\exp{(q_1\phi+q_2 \phi_2)}  =f   q_0+ g   q_1+h   q_2,\nonumber\\
\psi_1&=&\exp{(q_1\phi)}           =f_1 q_0+ g_1 q_1+h_1 q_2,\nonumber\\
\psi_2&=&\exp{(q_2\phi)}           =f_2 q_0+ h_2 q_1+g_2 q_2,\nonumber\\
\end{eqnarray}
Consequently, we can now introduce the conjugation operations for
these functions. For example, for $\psi_1$ we can get (zdesj):

\begin{equation}
\psi_1 {\bar \psi}_1 {\bar{\bar \psi}}_1 = \exp{( q_1\phi)}\exp{(j
\cdot q_1\phi)}\exp{(j^2 \cdot q_1\phi)}=q_0,
\end{equation}
which gives us the following link between   the functions,
$f,g,h$:
\begin{equation}
f_1^3+g_1^3+h_1^3-3 f_1g_1h_1=1.
\end{equation}
This surface (see figure \ref{fig:pic1s}) is a ternary analogue of
the $S^1$ circle and it is related with the ternary Abelian group,
$TU(1)$.

%\begin{figure}[htpb]
 % \begin{center}
 %   \mbox{\epsfig{file=pic1s.eps}}
 % \end{center}
%  \caption{The surface for $f_1^3+g_1^3+h_1^3-3 f_1g_1h_1=1$}
%  \label{fig:pig}
%\end{figure}

The Euler formula:

\begin{eqnarray}
z&=& \rho \exp {(\phi_1 q + \phi_2 q^2)}=\rho \exp {(\theta (q-q^2)+\phi (q+q^2)} \nonumber\\
&=& \rho ( c(\phi_1, \phi_2) + s(\phi_1,\phi_2) q + t(\phi_1,\phi_2) q^2),
\end{eqnarray}

where the Appel ternary trigonometric functions
\begin{eqnarray}
c&=& \frac{1}{3}( \exp {(\phi_1+\phi_2)}+    \exp {(j\phi_1+j^2\phi_2)}+    \exp {(j^2\phi_1+ j\phi_2)}) \nonumber\\
s&=& \frac{1}{3}( \exp {(\phi_1+\phi_2)}+ j^2\exp {(j\phi_1+j^2\phi_2)}+ j  \exp {(j^2\phi_1+j\phi_2)}) \nonumber\\
t&=& \frac{1}{3}( \exp {(\phi_1+\phi_2)}+ j  \exp {(j\phi_1+j^2\phi_2)}+ j^2\exp {(j^2\phi_1+j \phi_2)}) \nonumber\\
\end{eqnarray}

satisfy to the following equation:
\begin{eqnarray}
c^3+s^3+t^3-3cst=1.
\end{eqnarray}

There is also can be considered the ternary  logaritmic function \cite{LRV}:

\begin{eqnarray}
\ln z &=&(\ln z)_0+ (\ln z)_1 q + (\ln z)_2 q^2 \nonumber\\
      &=& (\ln \rho)_0+ \phi_1 q+\phi_2 q^2, \nonumber\\
\end{eqnarray}
where
\begin{eqnarray}
(\ln z)_0&=& \frac{1}{3} \ln ( x_0^3+x_1^3+x_2^3-3x_0x_1x_2) \nonumber\\
(\ln z)_1&=& \frac{1}{3} [\ln ( x_0+x_1+x_2)+ j^2 \ln ( x_0+jx_1+j^2x_2)+j\ln(x_0+j^2x_1+jx_2)] \nonumber\\
(\ln z)_2&=& \frac{1}{3} [\ln ( x_0+x_1+x_2)+ j   \ln ( x_0+jx_1+j^2x_2)+j^2\ln(x_0+j^2x_1+jx_2)] \nonumber\\
\end{eqnarray}

\noindent For further use, note that for elements $z,\tilde{z}$ and $\tilde{%
\tilde{z}}$ of the algebras\footnote{%
If we complexify the ternary complex numbers ${
T}_{3}C^{ C}={ T}_{3}
C\otimes_{ R}C$, the three above copies become
identical and $\ \ \tilde{}\ $ is an automorphism.} ${
T}_{3} C, \tilde{ T}_{3}C$ and
$\tilde{\tilde{ T}}_{3}C$ we have
$\tilde{z}+\tilde{\tilde{z}}=2 x_{0}-x_{1}q-x_{2}q^{2}
\in {T}_{3}\ C$, $%
\tilde{z}{\tilde{\tilde{z}}}%
=(x_{0}^{2}-x_{1}x_{2})+(x_{2}^{2}-x_{0}x_{1})q+(x_{1}^{2}-x_{2}x_{0})q^{2}%
\in {T}_{3}C$. We also have

\[
\begin{array}{lcll}
\parallel \ \parallel : & { T}_{3} C
\otimes \tilde{ T}_{3} C\otimes \tilde{%
\tilde{ T}}_{3} C & \rightarrow &  R\,, \\
& z\otimes \tilde{z}\otimes {\tilde{\tilde{z}}} & \mapsto &
\parallel
z\parallel ^{3}=z\tilde{z}{\tilde{\tilde{z}}}%
=x_{0}^{3}+x_{1}^{3}+x_{2}^{3}-3x_{0}x_{1}x_{2}
\end{array}
\]

. Thus, $\parallel
z\parallel =0 $  \textit{if and only if} $z$ belongs to $I_{1}$ or
to $I_{2}$. A ternary complex number is called non-singular if
$\parallel z\parallel \neq 0$.
>From now on we also denote $|z|$ the modulus of $z$.

It was proven in \cite{mc1} that any non-singular ternary complex number $%
z\in {T}_{3} C$ can be written in the ``polar
form'':

\begin{eqnarray}  \label{polar}
z= \rho e^{\varphi_1 q + \varphi_2 q^2}=\rho e^{\theta(q-q^2) +
\varphi(q+q^2)}
\end{eqnarray}

\noindent with $\rho = |z| =
\sqrt[3]{x_{0}^{3}+x_{1}^{3}+x_{2}^{3}-3x_{0}x_{1}x_{2}%
}\in   R,\theta \in \lbrack 0,2\pi/\sqrt{3} \lbrack ,
\varphi \in  R $. The combinations $q-q^{2}$ and $q+q^{2}$
generate in the ternary space
respectively compact and non-compact directions. Using
$q+q^{2}=2K_{0}+E_{0}$, we can rewrite in the form

\begin{eqnarray}  \label{mus}
z=\rho [m_0(\varphi_1,\varphi_2) + m_1(\varphi_1,\varphi_2)q +
m_2(\varphi_1,\varphi_2)q^2]
\end{eqnarray}

Since for the product of two ternary complex numbers we have
$\parallel zw\parallel =\parallel z\parallel \parallel w\parallel
$ the set of unimodular ternary complex numbers preserves the
cubic form \cite{mc1, int}.
The continuous group of symmetry of the cubic surface $%
x_{0}^{3}+x_{1}^{3}+x_{2}^{3}-3x_{0}x_{1}x_{2}=\rho ^{3}$ is isomorphic to $%
SO(2)\times SO(1,1)$. We denote the set of unimodular ternary
complex numbers or the ``ternary unit sphere'' as $ TU(1)=
\left\{ e^{(\theta +\varphi )q+(\varphi -\theta )q^{2}},\ 0\leq
\theta <2\pi/\sqrt{3} ,\varphi \in  R\right\} \sim
T S^{1}$.

>From the above figure one can see, that this surface approaches
asymptotically the plane $x_{0}+x_{1}+x_{2}=0$ and the line
 $x_{0}=x_{1}=x_{2}$ orthogonal to it. In $ T_{3} C$
they correspond to the ideals $I_{2}$ and $I_{1}$, respectively.
The latter line will be called the ``trisectrice''.

Let give the  Pithagore theorem through  the differential
2-forms.
One can construct the inner metric of this surface for the general
case $\rho \ne 0$. Introduce $a=x_0+x_1+x_2$ and parametrise a
point on the circle of radius $r$ around the trisectrice by its
polar coordinates $(r,\theta)$. The surface

\begin{equation}
x_0^3+x_1^3+x_2^3-3\,x_0x_1x_2=\rho^3
\end{equation}

in these coordinates
has the simple equation
\begin{equation}
ar^{2}=\rho ^{3}.
\end{equation}
 It can be shown that
for this cubic surface we can choose a parametrization,
$g(a, \theta): { R}^2 \rightarrow \Sigma $ for  a point $M(x_0,x_1,x_2)$:
\begin{eqnarray}
g(a,\theta)=(x_0(a, \theta),x_1(a, \theta),x_2(a, \theta) )
\end{eqnarray}
where
\begin{eqnarray}
x_0(a, \theta)&=&\frac{a}{3} -\frac{2}{3}\frac{\rho^{3/2}}{\sqrt{a}}
 \cos {\theta}, \nonumber\\
x_1(a, \theta)&=&\frac{a}{3} +\frac{1}{3}\frac{\rho^{3/2}}{\sqrt{a}}
( \cos {\theta}
+\sqrt{3}\sin \theta), \nonumber\\
x_2(a, \theta)&=&\frac{a}{3} +\frac{1}{3}\frac{\rho^{3/2}}{\sqrt{a}}
 (\cos {\theta} -\sqrt{3}\sin { \theta}).\nonumber\\
\end{eqnarray}

Now one can find the tangent vectors to the  surface
 $\Sigma \subset { R}^3 $
in the point $x_0(a,\theta),x_1(a,\theta),2_0(a,\theta) $
\begin{eqnarray}
\frac{\partial g}{\partial a}&=& (\frac{\partial x_0}{\partial a},
\frac{\partial x_1}{\partial a},\frac{\partial x_2}{\partial a}   ) \nonumber\\
\frac{\partial g}{\partial \theta}&=& (\frac{\partial x_0}{\partial \theta},
\frac{\partial x_1}{\partial \theta},\frac{\partial x_2}{\partial \theta}   ) \nonumber\\
\end{eqnarray}

or

\begin{eqnarray}
\frac{\partial x_0}{\partial a}&=& \frac{1}{3}+ \frac{1}{3}\frac{\rho^{3/2}}{a^{3/2}}
\cos \theta  \nonumber\\
\frac{\partial x_1}{\partial a}&=& \frac{1}{3}- \frac{1}{6}\frac{\rho^{3/2}}{a^{3/2}}
(\cos \theta +\sqrt{3} \sin \theta) \nonumber\\
\frac{\partial x_2}{\partial a}&=& \frac{1}{3}- \frac{1}{6}\frac{\rho^{3/2}}{a^{3/2}}
(\cos \theta - \sqrt{3} \sin \theta)  \nonumber\\
\end{eqnarray}

and

\begin{eqnarray}
\frac{\partial x_0}{\partial \theta}
&=& \frac{2}{3}\frac{\rho^{3/2}}{\sqrt{a}} \sin \theta  \nonumber\\
\frac{\partial x_1}{\partial \theta}
&=& \frac{1}{3}\frac{\rho^{3/2}}{\sqrt{a}} (-\sin \theta
+ \sqrt{3}\cos \theta)  \nonumber\\
\frac{\partial x_2}{\partial \theta}
&=& \frac{1}{3}\frac{\rho^{3/2}}{\sqrt{a}} (-\sin \theta
- \sqrt{3}\cos \theta)  \nonumber\\
\end{eqnarray}

These two tangent vectors allow to calculate the area of  the parallelogram based on them:

\begin{eqnarray}
J_{123}=
\left |
\begin{array}{ccc}

\frac{\partial x_0}{\partial a}
&\frac{\partial x_1}{\partial a}
&\frac{\partial x_2}{\partial a}  \\
\frac{\partial x_0}{\partial \theta}
&\frac{\partial x_1}{\partial \theta}
&\frac{\partial x_2}{\partial \theta}\\
 1  & 1  &  1\\
\end{array}
\right | .
\end{eqnarray}

Geometrically, the  differential forms $d x_0 \wedge  d x_1 $
$d x_1 \wedge  d x_2 $, $d x_2 \wedge  d x_0 $
are the areas of the parallelograms spanned by the vectors
$\frac{\partial g}{\partial a}$ and $\frac{\partial g}{\partial \theta}$
projected onto the $dx_0-dx_1$, $dx_1-dx_2$, $dx_2-dx_0$ planes, respectively.
This gives

\begin{eqnarray}
dx_k \wedge x_l = J_{kl}\, d a d \theta, \qquad  k,l =0,1,2,
\end{eqnarray}
where the Jacobians are
\begin{eqnarray}
J_{01}=
\left |
\begin{array}{ccc}
 \frac{\partial x_k}{\partial a}& \frac{\partial x_l}{\partial a}& 0 \\
 \frac{\partial x_k}{\partial \theta} & \frac{\partial x_l}{\partial \theta}& 0\\
0  &  0 &  1  \\
\end{array}
\right |
\end{eqnarray}

\begin{eqnarray}
J_{12}=
\left |
\begin{array}{ccc}
1 & 0 & 0 \\
 0 &\frac{\partial x_k}{\partial a}& \frac{\partial x_l}{\partial a}\\
 0 &\frac{\partial x_k}{\partial \theta} & \frac{\partial x_l}{\partial \theta}\\
\end{array}
\right |
\end{eqnarray}

\begin{eqnarray}
J_{20}=
\left |
\begin{array}{ccc}
 \frac{\partial x_k}{\partial a}& 0 & \frac{\partial x_l}{\partial a} \\
 0 & 1 & 0 \\
 \frac{\partial x_k}{\partial \theta} & 0 & \frac{\partial x_l}{\partial \theta}\\
\end{array}
\right |
\end{eqnarray}

Now one can see the geometrical meaning of $J_{01}$,$J_{12}$, $J_{20}$
and $J_{012}$ and to get the ternary analog of the Pithagorean theorem:

\begin{eqnarray}
 J_{01}^3+J_{12}^3+J_{20}^3- 3 J_{01}J_{12}J_{20}= J_{012}^3=\frac{1}{3 \sqrt{3}}
\frac{\rho^{6}}{a^{3}}.
\end{eqnarray}

From parallelogram Pithagore theorem one can easily come to the
tetrahedron Pithagore theorem:

\begin{eqnarray}
 S_A^3+S_B^3+S_B^3- 3 S_AS_BS_C= S_D^3,
\end{eqnarray}
where we have for $S_{...}$ four triangle faces of the tetrahedron.

The $TSO(2)\times TSO(1,1)$ group of transformations are generated
by the ternary-sine functions. In particular, in the special case
where $\varphi =0$
the transformation in the compact direction is a rotation to the angle $%
\sqrt{3}\theta $ and for $\theta =0$ we have the dilatation in the
non-compact direction

\[
\begin{array}{l}
\varphi =0:\left\{
\begin{array}{l}
x_{0}+x_{1}+x_{2}\rightarrow x_{0}+x_{1}+x_{2} \\
x_{0}+jx_{1}+j^{2}x_{2}\rightarrow e^{i\sqrt{3}\theta
}(x_{0}+jx_{1}+j^{2}x_{2})
\end{array}
\right. \,, \\
\\
\theta =0:\left\{
\begin{array}{l}
x_{0}+x_{1}+x_{2}\rightarrow e^{2\varphi }(x_{0}+x_{1}+x_{2}) \\
x_{0}+jx_{1}+j^{2}x_{2}\rightarrow e^{-\varphi
}(x_{0}+jx_{1}+j^{2}x_{2})
\end{array}
\right. \,.
\end{array}
\]

Let us consider now the discrete transformation preserving the modulus $%
\parallel z\parallel $ of non-singular ternary complex numbers:

\begin{eqnarray}  \label{duality}
z= \rho e^{\varphi_1 q + \varphi_2 q^2} \to \bar z = \frac{\tilde
z \tilde {\tilde z }}{\parallel z \parallel} =\rho e^{-\varphi_1 q
- \varphi_2 q^2}.
\end{eqnarray}

We  are going to investigate new aspects of the
ternary complex
analysis based on the ``complexification'' of ${R}^{3}$ space.
The use of the cyclic $\mathit{C}_{3}$ group for this purpose is
a natural generalization of the similar procedure for the
$\mathit{C}_{2}={Z}_{2}$ group in two dimensions. It is known that
the complexification of ${R}^{2}$ allows to introduce the new
geometrical objects - the Riemannian surfaces. The Riemannian
surfaces are defined as a pair $(M,C)$, where $M$ is a connected
two-dimensional manifold and $C$ is a complex structure on $M$.
Well-known examples of Riemann surfaces are the complex plane-
${C}$, Riemann sphere - ${CP}^{1}:C\cup {\inf }$ and complex tori-
${T}={C}/{\Gamma }$, $\Gamma :=n\lambda _{1}+n\lambda _{2}:n,m\in
{Z}$, $\lambda _{1,2}\in {C} $.

Let us introduce the complex valued functions $f(x_0,x_1)=a(x_0,x_1)+i b(x_0,x_1) $
 in an open subset $U\subset {C}$.

The harmonic functions $a(x_0,x_1)$ and $b(x_0,x_1)$ satisfy the Laplace
equations:

\begin{eqnarray}
&&\frac{\partial ^{2}a}{\partial z\partial \bar{z}}dz\wedge d\bar{z}=\frac{1%
}{2i}(\frac{\partial ^{2}a}{\partial x_0^{2}}+\frac{\partial
^{2}a}{\partial
x_1^{2}})dx\wedge dy=0\,,  \nonumber \\
&&\frac{\partial ^{2}b}{\partial z\partial \bar{z}}dz\wedge d\bar{z}=\frac{1%
}{2i}(\frac{\partial ^{2}b}{\partial x_0^{2}}+\frac{\partial
^{2}b}{\partial x_1^{2}})dx\wedge dy=0\,.
\end{eqnarray}
These equations are invariant under the $SO(2)$ transformations,
which is a
consequence of the symmetry of the $U(1)$ bilinear form
$\{z\bar{z}=(x_0+i x_1)(x_0-i x_1)
=1\}=S^{1} $ under the phase multiplication:
\[
z\rightarrow \exp \{i\alpha \}z,\qquad \bar{z}\rightarrow \exp \{-i\alpha \}%
\bar{z}\,.
\]

According to Dirac one can make the square root from Laplace
equation:

\begin{equation}
\sigma_1 \frac{\partial \psi}{\partial x_0}+ \sigma_2 \frac{\partial
\psi}{\partial x_1} =0,
\end{equation}

where a field $\psi$ is two-dimensional spinor
\begin{eqnarray}
\psi =
\left (
\begin{array}{c}
\phi_1 \\
\phi_2\\
\end{array}
\right )
\end{eqnarray}

and
where
\begin{eqnarray}
 \sigma_1=
 \left (
 \begin{array}{cc}
0 & 1 \\
1  & 0\\
\end{array}
\right ), \qquad
 \sigma_2=
 \left (
 \begin{array}{cc}
0 & -i \\
i & 0\\
\end{array}
\right )
\end{eqnarray}

 are the famous Pauli
matrices:
\begin{eqnarray}
 \sigma_m \sigma_n + \sigma_n \sigma_n=2 \delta_{mn}, \qquad m,n=1,2,
\end{eqnarray}
 which with $\sigma_3$ and$\sigma_0$ are
\begin{eqnarray}
\sigma_3&=&i \sigma_1 \sigma_2
=\left(
\begin{array}{cc}
1 & 0 \\
0 & -1 \\
\end{array}
\right ), \nonumber\\
\sigma_0&=& \sigma_i^2 =
\left(
\begin{array}{cc}
1 & 0 \\
0 & 1 \\
\end{array}
\right ), \qquad i=1,2,3. \nonumber\\
\end{eqnarray}

Thus on the complex plane  we have the following Dirac relation:

\begin{equation}
(\sigma_1 \frac{\partial }{\partial x_0}+ \sigma_2 \frac{\partial
}{\partial x_1})^2 =
\frac{\partial^2 }{\partial x_0^2}+ \frac{\partial^2
}{\partial x_1^2}
\end{equation}

Due to properties of all $\sigma_i$, $i=1,2,3$ matrices
the similar link remains valid in $D=3$:
\begin{equation}
(\sigma_1 \frac{\partial }{\partial x_0}+ \sigma_2 \frac{\partial
}{\partial x_1}+ \sigma_3 \frac{\partial
}{\partial x_2} )^2 =
\frac{\partial^2 }{\partial x_0^2}+ \frac{\partial^2
}{\partial x_1^2} + \frac{\partial^2
}{\partial x_2^2}.
\end{equation}

In $D=4$ one can get similar link if take into account the conjugation
properties of quaternions:
\begin{eqnarray}
&&(\sigma_0 \frac{\partial }{\partial x_0}+ i\sigma_1 \frac{\partial
}{\partial x_1}+ i\sigma_2 \frac{\partial
}{\partial x_2}+ i\sigma_3 \frac{\partial
}{\partial x_3} )\cdot
(\sigma_0 \frac{\partial }{\partial x_0}-i \sigma_1 \frac{\partial
}{\partial x_1}-i \sigma_2 \frac{\partial
}{\partial x_2}-i \sigma_3 \frac{\partial
}{\partial x_3} ) \nonumber\\
&=&
\frac{\partial^2 }{\partial x_0^2}+ \frac{\partial^2
}{\partial x_1^2} + \frac{\partial^2
}{\partial x_2^2}+\frac{\partial^2 }{\partial x_3^2}.
\end{eqnarray}

Note that  through the Pauli
matrices:
\begin{eqnarray}
\sigma_0, \,\,{\bf i} \sigma_1, {\bf i} \sigma_2, {\bf i} \sigma_3,
\end{eqnarray}

there is the matrix realization of quaternions
\begin{equation}
q=x_0e_0 +x_1e_1+x_2e_2+x_3e_3, \qquad q\in {H},
\end{equation}
which produce over ${ R}$  a 4-dimensional norm division algebra
where appears the third  imaginary unit
$e_3=e_1e_2 \equiv {\bf k}$.

The set of Pauli matrices produces the Clifford algebra
\begin{eqnarray}
&&\sigma_0 \nonumber\\
&&\sigma_1, \sigma_2 \nonumber\\
&&\sigma_1 \sigma_2 \nonumber\\
\end{eqnarray}
 and solution of linearized Dirac equation one should look for through the
 spinor fields:

\begin{equation}
\psi = \left (
\begin{array}{c}
\phi_1 \\
\phi_2 \\
\end{array}
\right )
\end{equation}

Thus in 2- dimensional space one can introduce the spin structure,
what was related to the complexification of ${ R}^2$. Dirac
 made the square root from the relativistic Klein-Gordon equation extending  the
binary Clifford algebra into four dimensional space-time:

\begin{equation}
\gamma_m \gamma_n+\gamma_n \gamma_m =2 g_{mn}, \qquad m,n=0,1,2,3.
\end{equation}
where

\begin{eqnarray}
\gamma_0&=&\sigma_1 \otimes \sigma_0
\nonumber\\
\gamma_1&=&\sigma_3 \otimes \sigma_0
\nonumber\\
\gamma_2&=&\sigma_2 \otimes \sigma_1
\nonumber\\
\gamma_3&=&\sigma_2 \otimes \sigma_3
\nonumber\\
\end{eqnarray}

In the relativistic  Dirac equation one should   consider already the bispinors  $
(\psi_1, \psi_2)$ which already have got  in addition to spin
structure a new geometrical structure related to the discovery
antiparticle states. each new structure will appear in
${ R}^{6,8,...}$ space.

Now consider the ${ C}_3$- holomorphicity.

Let us consider the function
\begin{eqnarray}
F(z,\tilde z, \tilde {\tilde z}) =f_0(x_0,x_1,x_2)
+f_1(x_0,x_1,x_2)q +f_2(x_0,x_1,x_2)q^2
\end{eqnarray}

For the $C_3$ holomorphicity we have two types:

\begin{itemize}
\item{1. For  the first type of  holomorphicity  function $F(
z,\tilde z,\tilde {\tilde z} )$ we have the following two
conditions:
\begin{eqnarray}
\frac{\partial F(z,\tilde z,\tilde {\tilde z})}{\partial \tilde
z}= \frac{\partial F(z,\tilde  z,\tilde {\tilde z})}{\partial
\tilde {\tilde z}}=0.
\end{eqnarray}
}\\
\item{2. For  the second  type of  holomorphicity  function $F(
z,\tilde z,\tilde {\tilde z})$ we can take  just one condition:
\begin{eqnarray}
\frac{\partial F(z,\tilde z,\tilde {\tilde z})}{\partial z}=0.
\end{eqnarray}
}\\
\end{itemize}

\begin{eqnarray}
\left (
\begin{array}{c}
\partial_{z_0}\\
\partial_{z_1}\\
\partial_{z_2}\\
\end{array}
\right )&=&||\frac{\partial x_p}{\partial z_r}|| \left (
\begin{array}{c}
\partial_0 \\
\partial_1 \\
\partial_2 \\
\end{array}
\right )
 = \left(
\begin{array}{ccc}
\frac{\partial x_0}{\partial z} &\frac{\partial x_1}{\partial z}
&\frac{\partial x_2}{\partial  z} \\
\frac{\partial x_0}{\partial  z_1} &\frac{\partial x_1}{\partial
z_1 } &\frac{\partial x_2}{\partial z_1}
\\\frac{\partial x_0}{\partial z_2} &\frac{\partial x_1}{\partial z_2}
&\frac{\partial x_2}{\partial z_2} \\
\end{array}
\right ) \left (
\begin{array}{c}
\partial_0 \\
\partial_1 \\
\partial_2 \\
\end{array}
\right ) \nonumber\\
 &=& \frac{1}{3}\left (
\begin{array}{ccc}
1    &     q^2    &     q    \\
1    & j^2 q^2    & j   q    \\
1    & j   q^2    & j^2 q    \\
\end{array}
\right ) \left (
\begin{array}{c}
\partial_0 \\
\partial_1 \\
\partial_2 \\
\end{array}
\right ) \nonumber\\
\end{eqnarray}

Here we used

\begin{eqnarray}
\left (
\begin{array}{c}
x_0\\
x_1\\
x_2\\
\end{array}
\right )= \frac{1}{3}\left (
\begin{array}{ccc}
1   &     1    &     1    \\
q^2 & j^2 q^2  & j   q^2  \\
q   & j   q    & j^2 q^2  \\
\end{array}
\right ) \left (
\begin{array}{c}
         z \\
\tilde z \\
\tilde {\tilde z} \\
\end{array}
\right )
\end{eqnarray}

More shortly:

\begin{eqnarray}
\partial_{z_p}=J_{pr}\partial_r
\end{eqnarray}

where
\begin{eqnarray}
J_{pr}=||\frac{\partial x_p}{\partial z_r}||= \frac{1}{3} \left (
\begin{array}{ccc}
1    &     q^2    &     q    \\
1    & j^2 q^2    & j   q    \\
1    & j   q^2    & j^2 q    \\
\end{array}
\right )
\end{eqnarray}

is Jacobian. We took some useful notations:

$\partial_{z_p} =\frac{\partial }{\partial z_p},$ and
$\partial_p=\frac{\partial }{\partial x_p}$, $z_1\equiv \tilde z$,
$z_2\equiv \tilde {\tilde z}$, $z_3 \equiv \tilde {\tilde{\tilde
z}}$, and $p,r=0,1,2$.

The inverse parities are the following:
\begin{eqnarray}
\left (
\begin{array}{c}
\partial_0 \\
\partial_1 \\
\partial_2 \\
\end{array}
\right )= \left (
\begin{array}{ccc}
1 &     1    &     1  \\
q & j   q    & j^2 q  \\
q^2 & j^2 q^2    & j   q^2  \\
\end{array}
\right ) \left (
\begin{array}{c}
\partial_{z_0} \\
\partial_{z_1} \\
\partial_{z_2} \\
\end{array}
\right )
\end{eqnarray}

As result, for all three derivatives  in we can give  the
following expressions:

\begin{eqnarray}
\partial_{z} F
&=& \frac{1}{3}(\partial_0 + q^2 \partial_1+q \partial_2)(f_0+q f_1+q^2 f_2)\nonumber\\
&=& \frac{1}{3}(\partial_0 f_0+\partial_1 f_1+\partial_2 f_2)    \nonumber\\
&+& \frac{1}{3}(\partial_2 f_0+\partial_0 f_1+\partial_1 f_2)q   \nonumber\\
&+& \frac{1}{3}(\partial_1 f_0+\partial_2 f_1+\partial_0 f_2)q^2,\nonumber\\
\end{eqnarray}

\begin{eqnarray}
\partial_{z_1} F&=&
   \frac{1}{3}( \partial_0 + j^2 q^2 \partial_1+ j q \partial_2)(f_0+q f_1+q^2+f_2)     \nonumber\\
&=&\frac{1}{3}(    \partial_0 f_0 + j^2\partial_1 f_1 + j   \partial_2 f_2)             \nonumber\\
&+&\frac{1}{3}(j   \partial_2 f_0 +    \partial_0 f_1 + j^2 \partial_1 f_2) q           \nonumber\\
&+&\frac{1}{3}(j^2 \partial_1 f_0 + j  \partial_2 f_1 +     \partial_0 f_2) q^2 ,       \nonumber\\
\end{eqnarray}

\begin{eqnarray}
\partial_{z_2} F
&=& \frac{1}{3}(\partial_0 + j q^2 \partial_1+ j^2 q \partial_2)(f_0+qf_1+q^2+f_2)\nonumber\\
&=& \frac{1}{3}(    \partial_0 f_0 + j  \partial_1 f_1 + j^2 \partial_2 f_2)      \nonumber\\
&+& \frac{1}{3}(j^2 \partial_2 f_0 +    \partial_0 f_1 + j   \partial_1 f_2) q    \nonumber\\
&+& \frac{1}{3}(j   \partial_1 f_0 + j^2\partial_2 f_1 +     \partial_0 f_2) q^2. \nonumber\\
\end{eqnarray}

The first type constraints $ \partial_{z_1} F=\partial_{z_2}=0$
give us the following differential equations:

\begin{eqnarray}
I. &&  \partial_0 f_0 + j^2\partial_1 f_1 + j   \partial_2 f_2 = 0,        \nonumber\\
II. &&j   \partial_2 f_0 +    \partial_0 f_1 + j^2 \partial_1 f_2 = 0,        \nonumber\\
III.&&j^2 \partial_1 f_0 + j  \partial_2 f_1 +     \partial_0 f_2 = 0         \nonumber\\
\end{eqnarray}

and

\begin{eqnarray}
IV.&&   \partial_0 f_0 + j  \partial_1 f_1 + j^2 \partial_2 f_2 = 0,    \nonumber\\
V.&& j^2 \partial_2 f_0 +    \partial_0 f_1 + j   \partial_1 f_2 = 0, \nonumber\\
VI.&&j   \partial_1 f_0 + j^2\partial_2 f_1 +     \partial_0 f_2)= 0. \nonumber\\
\end{eqnarray}

 Let solve the system of these  six  equations. For this let take
 the first equations, I and IV, from both system, multiply the equation I on the
 $j^2$
 and the equation IV on $j$:
\begin{eqnarray}
j^2 \partial_0 f_0 + j    \partial_1 f_1 +    \partial_2 f_2 &=& 0,        \nonumber\\
j   \partial_0 f_0 + j^2  \partial_1 f_1 +    \partial_2 f_2 &=& 0.    \nonumber\\
\end{eqnarray}
Having taken the difference of the equations one can get the
Cauchy-Riemann parity:

\begin{eqnarray}
\partial_0 f_0 =\partial_1 f_1
\end{eqnarray}

Similarly, one can get the full system of the linear differential
equations:

\begin{eqnarray}
\partial_0 f_0 &=&\partial_1 f_1=\partial_2 f_2 \nonumber\\
\partial_1 f_0 &=&\partial_2 f_1=\partial_0 f_2 \nonumber\\
\partial_2 f_0 &=&\partial_0 f_1=\partial_1 f_2 \nonumber\\
\end{eqnarray}

These equations give the definition of ternary harmonics
functions, the analogue of the Caushi-Riemann ( Darbu-Euleur)
definition for holomorphic  functions in the ordinary binary case.

From these equations one can get also that the three harmonics
functions, $f_0(x_0,x_1,x_2)$, $f_1(x_0,x_1,x_2)$, $f_2(x_0,x_1,x_2)$,
defined from
the holomorphic function
\begin{eqnarray}
F(z)=f_0(x_0,x_1,x_2) + q f_1(x_0,x_1,x_2)+ q^2 f_2(x_0,x_1,x_2)
\end{eqnarray}
are satisfied to the  cubic  Laplace equations:

\begin{eqnarray}
\partial_0^3 f_p +\partial_1^3 f_p +\partial_2^3 f_p
-3\partial_0\partial_1\partial_2 f_p=0, \quad p=0,1,2.
\end{eqnarray}

Let show  this for the harmonics function $f_0(x,y,u)$. For this
one should build the next combinations:

\begin{eqnarray}
\partial_0^3 f_0&=&\partial_0^2 \partial_1 f_1 =\partial_0^2
\partial_1 f_2
\nonumber\\
\partial_1^3 f_0&=&\partial_1^2 \partial_2 f_1 =\partial_1^2
\partial_0 f_2
\nonumber\\
\partial_2^3 f_0&=&\partial_2^2 \partial_0 f_1 =\partial_2^2
\partial_1 f_2
\nonumber\\
\end{eqnarray}
and
\begin{eqnarray}
\partial_0 \partial_1 \partial_2  f_0&=&\partial_2^2 \partial_0 f_1
\nonumber\\
\partial_0 \partial_1 \partial_2  f_0&=&\partial_1^2 \partial_2 f_1
\nonumber\\
\partial_0 \partial_1 \partial_2  f_0&=&\partial_0^2 \partial_1 f_1
\nonumber\\
\end{eqnarray}

Compare the two systems of differential equations for $f_0(x_0,x_1,x_2)$
one can get the ternary Laplace equation. Similarly, one can get
such equations for harmonics functions $f_1(x_0,x_1,x_2)$ and
$f_2(x_0,x_1,x_2)$.

Thus the  ternary holomorphic analysis in ${ R}^3$ leads to
ternary harmonic functions: $ f(z)=f_0(x_0,x_1,x_2)+q
f_1(x_0,x_1,x_2)+ q^2 f_2(x_0,x_1,x_2)$ which satified to cubic
differential equations:
\begin{equation}
\frac{\partial^3 f_i}{\partial x_0^3} + \frac{\partial^3
f_i}{\partial x_1^3}+ \frac{\partial^3 f_i}{\partial
x_2^3}-3\frac{\partial^3 f_i}{\partial x_0 \partial x_1 \partial
x_3}=0
\end{equation}

Let introduce the $3 \times 3$ matrices:

\begin{eqnarray}
Q_1 = \left (
\begin{array}{ccc}
0 & 1  & 0 \\
0 & 0  &  1 \\
1  & 0  &  0 \\
\end{array}
\right ), \qquad Q_2 = \left (
\begin{array}{ccc}
0 & 1  & 0 \\
0 & 0  &  j \\
j^2  & 0  &  0 \\
\end{array}
\right ), \qquad Q_3 = \left (
\begin{array}{ccc}
0 & 1  & 0 \\
0 & 0  &  j^2 \\
j  & 0  &  0 \\
\end{array}
\right )
\end{eqnarray}

These matrices satisfy to some  remarkable relations:
\begin{equation}
Q_aQ_bQ_c + Q_bQ_cQ_a+Q_cQ_aQ_b= 3 \eta_{abc} E_0
\end{equation}
with
\begin{eqnarray}
&&\eta_{111}=\eta_{222}=\eta_{333}=1 \nonumber\\
&&\eta_{123}=\eta_{231}=\eta_{312}=j \nonumber\\
&&\eta_{321}=\eta_{213}=\eta_{132}=j^2 \nonumber\\
\end{eqnarray}
where $j=\exp (2 \pi/3)$.

Using these matrices one can get the ternary Dirac equation:
\begin{eqnarray}
Q_1 \frac{\partial \Psi}{\partial x_0}+ Q_2 \frac{\partial
\Psi}{\partial x_1}+Q_3 \frac{\partial \Psi}{\partial x_2}=0,
\end{eqnarray}
where
\begin{eqnarray}
\Psi=(\psi_1,\psi_2,\psi_3),
\end{eqnarray}
is triplet of the wave functions, {\it i.e.} we introduced the
ternary spin structure in ${R}^3$. The next ternary
structures can appear in ${ R}^{6,9,12,...}$ spaces.

In order to diagonalize this equation we must act three times with
the same operator and we will get the cubic differential equation
satisfied by each component $\psi_p$, $ p=1,2,3$.

\newpage
\section{The  symmetry of the cubic forms}

The complex number theory is a seminal field in mathematics having
many applications to geometry, group theory, algebra and also to
the classical and quantum physics. Geometrically, it is based on
the complexification of the ${R}^{2}$ plane. The existence of
similar structures in higher dimensional spaces is interesting for
phenomenological applications.

It is  easily to check the following relation:
\begin{eqnarray}
\langle{\hat z}_1{\hat z}_2\rangle=\langle{\hat z}_1\rangle\langle{\hat z}_2\rangle,
\end{eqnarray}
which indicates  the group properties of the    ${ TC}$ numbers.
More exactly, the unit ${ TC}$ numbers produce the  Abelian ternary group.
According to the ternary analogue of the Euler formula, the following "unitary" ternary
$TU(1)$ group  can be constructed:

\begin{eqnarray}
U =\exp{(q \alpha+ q^2\beta) }
\end{eqnarray}

where $\alpha, \beta$ are the  group  parameters.
The 'unitarity" condition is:

\begin{eqnarray}
U \cdot \tilde U \cdot \tilde{\tilde U}=\hat 1,
\end{eqnarray}
where
\begin{eqnarray}
&&\tilde U =\exp{(jq \alpha+ j^2q^2\beta) }\nonumber\\
&&   \tilde {\tilde U} =\exp{(j^2q \alpha+ jq^2\beta) },\nonumber\\
\end{eqnarray}

Similarly to the binary case, when for the $U(1)$ Abelian group one
can find the form of $SO(2)$ group, there also  exists
such a  correspondence. For simplicity,  take $\beta =0$. Then

\begin{eqnarray}
U \rightarrow O&=&
\left (
\begin{array}{ccc}
c & s & t \\
t & c & s\\
s & t & c \\
\end{array}
\right ) \nonumber\\
\tilde U \rightarrow \tilde O&=&
\left (
\begin{array}{ccc}
c & js & j^2t \\
j^2t & c & js\\
js & j^2t & c \\
\end{array}
\right ) \nonumber\\
\tilde {\tilde U} \rightarrow
\tilde {\tilde O}
&=&
\left (
\begin{array}{ccc}
c & j^2s & jt \\
jt & c & j^2s\\
j^2s & jt & c \\
\end{array}
\right ) \nonumber\\
\end{eqnarray}
where

\begin{eqnarray}
O \cdot \tilde O \cdot \tilde {\tilde O}=
c^3+s^3+t^3-3cst \cdot
\left (
\begin{array}{ccc}
1 & 0  & 0 \\
0 & 1  & 0\\
0 & 0  & 1 \\
\end{array}
\right ).
\end{eqnarray}

Let us find

\begin{eqnarray}
( x^{\prime}, y^{\prime},u^{\prime})^t = O_{Si} \cdot (x,y,u)^t.
\end{eqnarray}

where
\begin{eqnarray}
O_{S}= \exp\{\alpha q_1+\beta q_1^2\}=O_{S1}O_{S2}
\end{eqnarray}
and
\begin{eqnarray}
O_{S1}= \exp\{\alpha q_1\}, \qquad O_{S2}= \exp\{\beta q_1^2\},
\end{eqnarray}
respectively.

The generators $q_1$ and $q_1^2$ can be represented in the matrix
form:
\begin{eqnarray}
 q_1=
\left (
\begin{array}{ccc}
    0  &     1   &   0      \\
    0  &     0   &   1       \\
    1  &     0   &   0        \\
\end{array}
\right), \qquad
 q_1^2=q_4=
\left (
\begin{array}{ccc}
    0  &     0   &   1      \\
    1  &     0   &   0       \\
    0  &     1   &   0        \\
\end{array}
\right)
\end{eqnarray}

Let find the eigenvalues
\begin{eqnarray}
 det \{\alpha q_1 +\beta q_1^2- \lambda E\}=
det \left (
\begin{array}{ccc}
    -\lambda   &   \alpha   &   \beta     \\
      \beta    &  -\lambda  &   \alpha    \\
     \alpha    &   \beta    &  -\lambda   \\
\end{array}
\right)
=-\lambda ^3 +\alpha^3+\beta^3-3 \lambda \alpha \beta=0 \nonumber\\
\end{eqnarray}
So, we have the following three eigenvalues:
\begin{eqnarray}
\lambda_1=   \alpha +     \beta,  \qquad \lambda_2=j  \alpha + j^2
\beta,  \qquad \lambda_3=j^2\alpha + j   \beta .
\end{eqnarray}
\begin{eqnarray}
O_{S}= SS^{-1}\exp\{\alpha q_1+\beta q_1^2\} SS^{-1}= S\exp\{
S^{-1}(\alpha q_1+\beta q_1^2) S\} S^{-1},
\end{eqnarray}
\begin{eqnarray}
 S=\frac{1}{\sqrt{3}}
\left (
\begin{array}{ccc}
    1  &     1     &   1      \\
    1  &     j     &   j^2       \\
    1  &     j^2   &   j        \\
\end{array}
\right), \qquad S^{-1}=S^+=\frac{1}{\sqrt{3}} \left (
\begin{array}{ccc}
    1  &     1     &   1      \\
    1  &     j^2   &   j       \\
    1  &     j     &   j^2        \\
\end{array}
\right),
\end{eqnarray}

\begin{eqnarray}
&&S\exp\{\alpha S^{-1}(\alpha q_1+\beta q_1^2) S\} S^{-1}=S\exp\{
\left (
\begin{array}{ccc}
    \alpha +     \beta  &     0                  &   0                \\
    0                   &  j  \alpha + j^2 \beta &   0 \\
    0                   &     0                  &  j^2\alpha + j\beta \\
\end{array}
\right )\} S^{-1}
\nonumber\\
&=& S\exp\{ \alpha \left (
\begin{array}{ccc}
    1     &  0  &   0 \\
    0     &  j  &   0  \\
    0     &  0  &  j^2 \\
\end{array}
\right )\} S^{-1} S\exp\{ \beta \left (
\begin{array}{ccc}
    1     &  0   &   0 \\
    0     &  j^2 &   0 \\
    0     &  0   &   j \\
\end{array}
\right )\} S^{-1}
\nonumber\\
&&= S\left (
\begin{array}{ccc}
  c_1(\alpha)+s_1(\alpha)+t_1(\alpha)&     0      &   0      \\
  0          & c_1(\alpha)+ js_1(\alpha)+j^2t_1(\alpha) &   0       \\
  0     &     0      &  c_1(\alpha)+j^2s_1(\alpha)+jt_1(\alpha)         \\
\end{array}
\right ) S^{-1} \nonumber \\
&&\cdot S\left (
\begin{array}{ccc}
  c_2(\beta)+s_2(\beta)+t_2(\beta)&     0      &   0      \\
  0          & c_2(\beta)+ j^2s_2(\beta)+jt_2(\beta) &   0       \\
  0     &     0      &  c_2(\beta)+js_2(\beta)+j^2t_2(\beta)         \\
\end{array}
\right ) S^{-1} \nonumber\\
&&= \left (
\begin{array}{ccc}
    c_1(\alpha)   &     s_1(\alpha)    &   t_1 (\alpha)        \\
    t_1(\alpha)   &     c_1(\alpha)    &   s_1 (\alpha)        \\
    s_1(\alpha)   &     t_1 (\alpha)   &   c_1(\alpha)         \\
\end{array}
\right ) \left (
\begin{array}{ccc}
    c_2(\beta)  &     t_2(\beta)   &   s_2(\beta)      \\
    s_2(\beta)  &     c_2(\beta)   &   t_2(\beta)        \\
    t_2(\beta)  &     s_2(\beta)   &   c_2(\beta)        \\
\end{array}
\right )
\nonumber\\
\end{eqnarray}

Let us consider two limit cases:
\begin{itemize}
\item{1 case:  $\alpha=-\beta$} \\
\item{2 case:  $\alpha= \beta$} \\
\end{itemize}

The case 1 is related to the binary orthogonal symmetry of the
cubic surface and cubic forms. This orthogonal symmetry is in the
plane which ortoghonal to the direction of `` trisectriss''.

\begin{eqnarray}
&&O_S= \exp \{\alpha (q_1-q_1^2)\}=
S\exp\{S^{-1}(\alpha(q_1-q_1^2)) S\} S^{-1} \nonumber\\
&&=S\exp\{ \left (
\begin{array}{ccc}
    0       &     0              &   0                \\
    0       &    \alpha (j- j^2) &   0                \\
    0       &     0              &   \alpha (j^2-j)   \\
\end{array}
\right )\} S^{-1}  \nonumber\\
&& =S \left (
\begin{array}{ccc}
    1       &     0                      &   0                        \\
    0       &    \exp\{\alpha (j- j^2)\} &   0                        \\
    0       &     0                      &   \exp\{\alpha (j^2-j)\}   \\
\end{array}
\right )\} S^{-1}  \nonumber\\
&&=\frac{1}{3}\left (
\begin{array}{ccc}
 1+   e^{\{i\phi\}}+   e^{\{-i\phi\}}
&1+j^2e^{\{i\phi\}}+j  e^{\{-i\phi\}}
&1+j  e^{\{i\phi\}}+j^2e^{\{-i\phi\}}    \\
 1+j  e^{\{i\phi\}}+j^2e^{\{-i\phi\}}
&1+   e^{\{i\phi\}}+   e^{\{-i\phi\}}
&1+j^2e^{\{i\phi\}}+   e^{\{-i\phi\}}    \\
 1+j^2e^{\{i\phi\}}+j  e^{\{-i\phi\}}
&1+j  e^{\{i\phi\}}+j^2e^{\{-i\phi\}}
&1+   e^{\{i\phi\}}+   e^{\{-i\phi\}}    \\
\end{array}
\right )  \nonumber\\
\end{eqnarray}
where $j-j^2=\sqrt{3}{\bf i}$, $\phi=\sqrt{3}\alpha$.

Thus
\begin{eqnarray}
O_S=\left (
\begin{array}{ccc}
 c_0& s_0 &t_0  \\
 t_0& c_0 &s_0  \\
 s_0& t_0 &c_0  \\
\end{array}
\right ),
\end{eqnarray}            \nonumber\\
where we have the particular choice for the functions, $c$, $s$,
$t$:
\begin{eqnarray}
&&c_0=\frac{1}{3}( 1+   e^{\{i\phi\}}+   e^{\{-i\phi\}})
=\frac{1}{3}(1+2 cos (\phi)) \nonumber\\
&&s_0=\frac{1}{3}(1+j^2e^{\{i\phi\}}+j  e^{\{-i\phi\}})
 =\frac{1}{3}(1+2 cos (\phi +\frac{2\pi}{3})), \nonumber\\
&&t_0=\frac{1}{3}(1+j  e^{\{i\phi\}}+j^2e^{\{-i\phi\}})
 =(\frac{1}{3}1+2 cos (\phi - \frac{2\pi}{3})). \nonumber\\
\end{eqnarray}

One can check that $c_0^3+s_0^3+t_0^3-3c_0s_0t_0=1$. But these
transformations are also binary orthogonal transformations. It
means that the matrices

\begin{eqnarray}
O=  \left (
\begin{array}{ccc}
    c_0  &     s_0   &  t_0     \\
    t_0  &     c_0   &  s_0     \\
    s_0  &     t_0   &  c_0         \\
\end{array}
\right ),
\end{eqnarray}
and
\begin{eqnarray}
O^t=  \left (
\begin{array}{ccc}
    c_0  &     t_0   &  s_0         \\
    s_0  &     c_0   &  t_0         \\
    t_0  &     s_0   &  c_0         \\
\end{array}
\right ),
\end{eqnarray}
satisfy to condition $O O^t=O^tO=1$, what is equivalent to the
additional  two equations:
\begin{eqnarray}
&& c_0^2+s_0^2+t_0^2=1 \nonumber\\
&& c_0s_0+s_0t_0+t_0c_0=0,  \nonumber\\
\end{eqnarray}
what in our case can be easily checked. Thus in the case 1 the
ternary symmetry coincides with the orthogonal binary symmetry
$SO(2)$.

In the case 2, $\alpha=\beta $, the ternary symmetry coincides
with the other binary symmetry.

\begin{eqnarray}
&&O_S= \exp \{\alpha (q_1+q_1^2)\}=
S\exp\{S^{-1}(\alpha(q_1+q_1^2)) S\} S^{-1} \nonumber\\
&&=S\exp\{ \alpha \left (
\begin{array}{ccc}
    2       &     0              &   0                \\
    0       &    -1              &   0                \\
    0       &     0              &  -1   \\
\end{array}
\right )\} S^{-1}  \nonumber\\
&& =S \left (
\begin{array}{ccc}
   \exp \{2 \alpha\}       &     0                      &   0                        \\
    0       &    \exp\{-\alpha\} &   0                        \\
    0       &     0                      &   \exp\{-\alpha\}   \\
\end{array}
\right ) S^{-1}  \nonumber\\
&&=\frac{1}{3}\left (
\begin{array}{ccc}
  e^{\{2\alpha\}}+  2 e^{\{-\alpha\}}
& e^{\{2\alpha\}}-    e^{\{-\alpha\}}
& e^{\{2\alpha\}}-    e^{\{-\alpha\}}  \\
  e^{\{2\alpha\}}-    e^{\{-\alpha\}}
& e^{\{2\alpha\}}+  2 e^{\{-\alpha\}}
& e^{\{2\alpha\}}-    e^{\{-\alpha\}}  \\
  e^{\{2\alpha\}}-    e^{\{-\alpha\}}
& e^{\{2\alpha\}}-    e^{\{-\alpha\}}
& e^{\{2\alpha\}}+  2 e^{\{-\alpha\}}  \\
\end{array}
\right ) . \nonumber\\
\end{eqnarray}

In this case the operator $O_S$ can be repesented in the following
more simpler form, {\it i.e.}

\begin{eqnarray}
O=  \left (
\begin{array}{ccc}
    c_{+}  &   s_{+}   &  s_{+}   \\
    s_{+}  &   c_{+}    &  s_{+}     \\
    s_{+}  &   s_{+}    &  c_{+}        \\
\end{array}
\right ),
\end{eqnarray}
where $ c_{+}= \frac{1}{3}(e^{\{2\alpha\}}+  2 e^{\{-\alpha\}} )$,
      $ s_{+}= \frac{1}{3}(e^{\{2\alpha\}}-    e^{\{-\alpha\}} )$
and the cubic equation reduces to the next form:
\begin{eqnarray}
 c_{+}^3+ s_{+}^3+ t_{+}^3-3 c_{+} s_{+} t_{+}=( c_{+}- s_{+})^2( c_{+}+2 s_{+})=1.
\end{eqnarray}

Thus, the two parametric ternary $TSO(2)$ group reduces exactly to
two known binary symmetries, $\alpha=-\beta$ and  $\alpha=\beta$,
but for the general case, it produces the new symmetry, in which
these two binary symmetry are unified by non-trivial way, ( it is
not product!)

Let us go further to study so me properties...

\begin{eqnarray}
\tilde O_{S1}&=& \exp\{j\alpha q_1\}=SS^{-1}\exp\{j\alpha q_1\}S^{-1}S\nonumber\\
&=&S \exp\{j\alpha (S^{-1}q_1S)\}=\exp\{j\alpha q_7\}S^{-1} \nonumber\\
&=&S[ \sum_{k=0}\frac{(j \alpha) ^{3k}}{3k!}+ q_7\sum_{k=0}\frac{
(j \alpha) ^{3k+1}}{(3k+1)!} +
q_7^2\sum_{k=0}\frac{(j \alpha) ^{3k+2}}{(3k+2)!}]S^{-1}, \nonumber\\
\end{eqnarray}
where
\begin{eqnarray}
q_7=S^{-1}q_1S=\left (
\begin{array}{ccc}
    1 & 0  &   0      \\
    0 & j  &   0      \\
    0 & 0  &   j^2    \\
\end{array}
\right).
\end{eqnarray}
Then one can get

\begin{eqnarray}
\tilde O_{S1} &=& S[\left (
\begin{array}{ccc}
    c & 0  &   0     \\
    0 & c  &   0     \\
    0 & 0  &   c     \\
\end{array}
\right )+ \left (
\begin{array}{ccc}
    js & 0     &   0     \\
    0  & j^2s  &   0     \\
    0  & 0     &   s    \\
\end{array}
\right ) +\left (
\begin{array}{ccc}
    j^2t & 0  &   0     \\
    0    & jt &   0     \\
    0    & 0  &   t     \\
\end{array}
\right )]
 S^{-1} \nonumber\\
&=& S\left (
\begin{array}{ccc}
    c+js+j^2t & 0          &   0        \\
    0         & c+j^2s+jt  &   0         \\
    0         & 0          &   c+s+t     \\
\end{array}
\right ) S^{-1}= \left (
\begin{array}{ccc}
       c &  j  s   &   j^2t        \\
    j^2t &     c   &   j  s         \\
    j  s &  j^2t   &      c           \\
\end{array}
\right )
\nonumber\\
\end{eqnarray}

So we have
\begin{eqnarray}
\tilde O_{S1}&=& \exp\{j\alpha q_1\} =\left (
\begin{array}{ccc}
       c &  j  s   &   j^2t        \\
    j^2t &     c   &   j  s         \\
    j  s &  j^2t   &      c           \\
\end{array}
\right )
\end{eqnarray}

Similarly,
\begin{eqnarray}
\tilde {\tilde O}_{S1}&=& \exp\{j^2 \alpha q_1\} =\left (
\begin{array}{ccc}
        c &  j^2 s &   j  t        \\
        j &      c &   j^2s        \\
    j^2 s &  j   t &      c        \\
\end{array}
\right )
\end{eqnarray}

One can easily to check that

\begin{eqnarray}
O_{S1}\tilde O_{S1}\tilde {\tilde O}_{S1}&=& \exp\{\alpha q_1\}
\exp\{\alpha \tilde q_1\}
\exp\{\alpha \tilde {\tilde q}_1\} \nonumber\\
&=& \exp\{ \alpha q_1\}\exp\{j \alpha q_1\}\exp\{j^2 \alpha q_1\}
=(c^3+s^3+t^3-3cst) \left (
\begin{array}{ccc}
        1 &  0  &   0        \\
        0 &  1  &   0        \\
        0 &  0  &   1        \\
\end{array}
\right ). \nonumber\\
\end{eqnarray}

One can check that   the ternary orthogonal transformations in the
following form:
\begin{eqnarray}
O_S=  \left (
\begin{array}{ccc}
    c  &     s   &  t     \\
    t  &     c   &  s     \\
    s  &     t   &  c         \\
\end{array}
\right ),
\end{eqnarray}
with  $c^3+s^3+t^3 -3 c s t =1$ conserve the cubic forms, {\it
i.e.}
\begin{eqnarray}
(x_0^\prime)^3+(x_1^\prime)^3+(x_2^\prime)^3
-3(x_0^\prime)(x_1^\prime)(x_2^\prime)=x_0^3+x_1^3+x_2^3-3x_0x_1x_2
\end{eqnarray}
or the cubic Laplace equations:
\begin{eqnarray}
&&\frac{\partial^3 a}{\partial x_0^3}+\frac{\partial^3 a}{\partial
x_1^3}+\frac{\partial^3 a}{\partial x_2^3} -3 \frac{\partial^3
a}{\partial x_0 \partial x_1 \partial x_2}=0,\nonumber\\
&&\frac{\partial^3 b}{\partial x_0^3}+\frac{\partial^3 b}{\partial
x_1^3}+\frac{\partial^3 b}{\partial x_2^3} -3
\frac{\partial^3 c}{\partial x_0 \partial x_1 \partial x_2}=0,\nonumber\\
&&\frac{\partial^3 c}{\partial x_0^3}+\frac{\partial^3 c}{\partial
x_1^3}+\frac{\partial^3 c}{\partial x_2^3} -3 \frac{\partial^3
c}{\partial x_0 \partial x_1 \partial x_2}=0,\nonumber\\
\end{eqnarray}

%Also, one can see that the ternary complex form also are invariant,
%{\it i.e.}
%\begin{eqnarray}
% x^{\prime} \tilde x^{\prime} \tilde {tilde x}^{\prime}
%+y^{\prime} \tilde y^{\prime} \tilde {tilde y}^{\prime}
%+u^{\prime} \tilde u^{\prime} \tilde {tilde u}^{\prime)
%-x^{\prime}  y^{\prime}  u^{\prime}
%-\tilde x^{\prime} \tilde y^{\prime} tilde u^{\prime}
%-x^{\prime} \tilde y^{\prime} \tilde {tilde u}^{\prime}
%=
%  x \tilde x \tilde {\tilde x}
%+ y \tilde y \tilde {\tilde y}
%+ u \tilde u \tilde {\tilde u}
%- x  y  u
%- \tilde x \tilde y \tilde u
%- \tilde {\tilde x} \tilde {\tilde y} \tilde { \tilde u}
%\end{eqnarray}
\section{Quaternary $C_4$- complex numbers}
Consider the quaternary  complex numbers

\begin{eqnarray}
z=x_0q_0 +x_1 q+x_2q^2+x_3q^3
\end{eqnarray}
where we can consider two cases:

\begin{eqnarray}
A: \,q^4=q_0=1
\end{eqnarray}
or

\begin{eqnarray}
B:\, q^4=-q_0=-1.
\end{eqnarray}
Let define the conjugation operation of a new complex number:

\begin{eqnarray}
 \tilde q_0=q_0=1,  \qquad  \tilde q= j q, \qquad where \qquad j^{4}=1,
\end{eqnarray}
namely
\begin{eqnarray}
j =\exp{i \pi/2}.
\end{eqnarray}

Now one can  calculate the norm of this complex number:
\begin{eqnarray}
 z \tilde z \tilde{\tilde  z} \tilde {\tilde {\tilde z}}=1,
\end{eqnarray}
where

\begin{eqnarray}
\begin{array}{ccccccccc}
                z
&=&x_0q_0 &+&x_1 q &+&x_2q^2
&+&x_3q^3                                  \\
\tilde         {z} &=& x_0 q _0 &+& x_1 \tilde {q} &+& x_2 \tilde
{q^2}
&+& x_3 \tilde {q^3}                       \\
\tilde {\tilde         {z}} &=& x_0 q_0 &+& x_1 \tilde {\tilde
{q}} &+& x_2 \tilde {\tilde {q^2}}
&+& x_3 \tilde {\tilde {q^3}}              \\
\tilde {\tilde {\tilde         {z}   }} &=& x_0 q_0 &+& x_1 \tilde
{\tilde {\tilde {q}   }} &+& x_2 \tilde {\tilde {\tilde {q^2} }}
&+& x_3 \tilde {\tilde {\tilde {q^3} }},    \\
\end{array}
\end{eqnarray}
or

\begin{eqnarray}
\begin{array}{ccccccccc}
z&=&x_0q_0 &+&x_1 q&+&x_2q^2&+&x_3q^3 \\
\tilde z&=&x_0q_0  &+& j x_1  q &+& j^2 x_2q^2 & +&j^3 x_3q^3 \\
\tilde {\tilde z}&=&x_0q_0  &+& j^2 x_1 q &+& j^4 x_2q^2 &+& j^6 x_3q^3 \\
\tilde {\tilde {\tilde z}}&=&x_0q_0  &+&j^3 x_1 q &+& j^6x_2q^2 &+&j^9 x_3q^3 \\
\end{array}
\end{eqnarray}

or

\begin{eqnarray}
\begin{array}{ccccccccc}
z&=&x_0q_0 &+&x_1 q&+&x_2q^2&+&x_3q^3 \\
\tilde z&=&x_0q_0  &+& i x_1  q &-& x_2q^2 &-&i x_3q^3 \\
\tilde {\tilde z}&=&x_0q_0  &-& x_1 q &+& x_2q^2 &-& x_3q^3 \\
\tilde {\tilde {\tilde z}}&=&x_0q_0  &-&i x_1 q &-& x_2q^2 &+&i x_3q^3 \\
\end{array}
\end{eqnarray}

We used the following relations:

\begin{eqnarray}
\begin{array}{|ccccc|ccccc|ccccc|}
                                                               \hline
\tilde {q}  &=& i q   &=& i q                                    &
\tilde {\tilde {q}}&=&i^2q&=&-q                          & \tilde
{\tilde {\tilde {q}}}&=& i^3 q&=&-iq              \\   \hline
 \tilde {q^2}&=& i^2 q &=& - q^2 &
 \tilde {\tilde {q^2}}&=&i^4q^2&=&q^2                     &
\tilde {\tilde {\tilde {q^2}}}&=&i^6 q^2&=&-q^2          \\
\hline \tilde {q^3}&=& i^3 q &=&- i q^3
& \tilde {\tilde {q^3}}&=&i^6 q^3&=&-q^3                   &
\tilde {\tilde {\tilde {q^3}}}&=&i^9 q^3&=&-i q^3        \\
\hline
\end{array}
\end{eqnarray}
To find the equation of the surface

\begin{eqnarray}
 z \tilde z \tilde{\tilde  z} \tilde {\tilde {\tilde z}}=1,
\end{eqnarray}
we should take into account the following identities:
\begin{eqnarray}
1+j+j^2+j^3=0, \qquad 1+j^2=0, \qquad j+j^3=0.
\end{eqnarray}

In the case $A$, $q^4=1$ the unit quaternary complex numbers
determine the following surface:
\begin{eqnarray}
z \tilde z \tilde{\tilde  z} \tilde {\tilde {\tilde z}} &=&
x_0^4-x_1^4+x_2^4-x_3^4 -2x_0^2x_2^2+2x_1^2x_3^2  \nonumber\\
&-&4x_0^2x_1x_3+4x_1^2x_0x_2-4x_2^2x_1x_3+4x_3^2x_0x_2\nonumber\\
&=&[x_0^2+x_2^2-2x_1x_3]^2-[x_1^2+x_3^2-2x_0x_2]^2 \nonumber\\
&=&[x_0^2+x_1^2+x_2^2+x_3^2-2x_1x_3-2x_0x_2]
[x_0^2-x_1^2+x_2^2-x_3^2-2x_1x_3+2x_0x_2=
\nonumber\\
&=&[(x_0-x_2)^2+(x_1-x_3)^2][(x_0^+x_2)^2-(x_1+x_3)^2]\nonumber\\
&=&(x_0+x_1+x_2+x_3)(x_0+x_2-x_1-x_3)[(x_0-x_2)^2+(x_1-x_3)^2]
 =1,
\end{eqnarray}
In the case $B$,  $q^4=-1$ one can get:

\begin{eqnarray}
z \tilde z \tilde{\tilde  z} \tilde {\tilde {\tilde z}} &=&
x_0^4+x_1^4+x_2^4+x_3^4+2x_0^2x_2^2+2x_1^2x_3^2  \nonumber\\
&+&4x_0^2x_1x_3-4x_1^2x_0x_2-4x_2^2x_1x_3+4x_3^2x_0x_2\nonumber\\
&=&[x_3^2-x_1^2+2x_0x_2]^2+[x_0^2-x_2^2+2x_1x_3]^2 =1,
\end{eqnarray}

For illustration consider the $Z_4$- holomorphicity for the case
A.

Let us consider the function
\begin{eqnarray}
&&F(z,\tilde z, \tilde {\tilde z}, \tilde {\tilde {\tilde z}})
\nonumber\\ &=&f_0(x_0,x_1,x_2,x_3)+f_1(x_0,x_1,x_2,x_3 )q
+f_2(x_0,x_1,x_2,x_3)q^2+f_3( x_0,x_1,x_2,x_3 )q^3 \nonumber\\
\end{eqnarray}
and her first derivatives:

\begin{eqnarray}
\partial_z F
&=& \frac{1}{4} \partial_0 F +\frac{1}{4}q^3 \partial_1 F
+\frac{1}{4} q^2\partial_2 F
+\frac{1}{4}q \partial_3 F \nonumber\\
\partial_{\tilde z} F
 &=& \frac{1}{4} \partial_0 F -\frac{i}{4} q^3 \partial_1 F
-\frac{1}{4}q^2 \partial_2 F +\frac{i}{4} q\partial_3 F \nonumber\\
\partial_{\tilde {\tilde z}} F
&=& \frac{1}{4} \partial_0 F -\frac{1}{4} q^3 \partial_1 F
+\frac{1}{4}q^2 \partial_2 F -\frac{1}{4} q\partial_3 F \nonumber\\
\partial_{\tilde {\tilde {\tilde z}}} F
&=& \frac{1}{4} \partial_0 F +\frac{i}{4} q^3 \partial_1 F
-\frac{1}{4}q^2 \partial_2 F -\frac{i}{4} q\partial_3 F \nonumber\\
\end{eqnarray}
where we used

\begin{eqnarray}
\left (
\begin{array}{c}
\partial_z \\
\partial_{\tilde z} \\
\partial_{\tilde {\tilde z}}\\
\partial_{ \tilde {\tilde {\tilde z}}}\\
\end{array}
\right ) = \frac{1}{4} \left (
\begin{array}{cccc}
1 &   q^3  &   q^2 &   q \\
1 &-i q^3  &  -q^2 &  iq \\
1 &-  q^3  &   q^2 &  -q \\
1 &i  q^3  &  -q^2 & -iq \\
\end{array}
\right )
\left(
\begin{array}{c}
\partial_0 \\
\partial_1 \\
\partial_2 \\
\partial_3\\
\end{array}
\right )
\end{eqnarray}
where

$\partial_{z_p} =\frac{\partial }{\partial z_p},$ and
$\partial_p=\frac{\partial }{\partial x_p}$ $p=0,1,2,3$,
$z_1\equiv \tilde z$, $z_2\equiv \tilde {\tilde z}$, $z_3 \equiv
\tilde {\tilde{\tilde z}}$.

\begin{eqnarray}
\partial_z F&=&
\frac{1}{4}(\partial_0 f_0+\partial_1 f_1
+\partial_2f_2+\partial_3 f_3)\nonumber\\
&+&\frac{1}{4}(\partial_0 f_1+\partial_1 f_2
+\partial_2f_3+\partial_3 f_0)q\nonumber\\
&+&\frac{1}{4}(\partial_0 f_2+\partial_1 f_3
+\partial_2f_0+\partial_3 f_1)q^2\nonumber\\
&+&\frac{1}{4}(\partial_0 f_3+\partial_1 f_0
+\partial_2f_1+\partial_3 f_2)q^3\nonumber\\
\end{eqnarray}

\begin{eqnarray}
\partial_{z_1} F&=&
\frac{1}{4}(\partial_0 f_0-i\partial_1 f_1
-\partial_2f_2+i\partial_3 f_3)\nonumber\\
&+&\frac{1}{4}(\partial_0 f_1-i\partial_1 f_2
-\partial_2f_3+i\partial_3 f_0)q\nonumber\\
&+&\frac{1}{4}(\partial_0 f_2-i\partial_1 f_3
-\partial_2f_0+i\partial_3 f_1)q^2\nonumber\\
&+&\frac{1}{4}(\partial_0 f_3-i\partial_1 f_0
-\partial_2f_1+i\partial_3 f_2)q^3\nonumber\\
\end{eqnarray}

\begin{eqnarray}
\partial_{z_2} F
&=&
\frac{1}{4}(\partial_0 f_0-\partial_1 f_1
+\partial_2f_2-\partial_3 f_3)\nonumber\\
&+&\frac{1}{4}(\partial_0 f_1-\partial_1 f_2
+\partial_2f_3-\partial_3 f_0)q\nonumber\\
&+&\frac{1}{4}(\partial_0 f_2-\partial_1 f_3
+\partial_2f_0-\partial_3 f_1)q^2\nonumber\\
&+&\frac{1}{4}(\partial_0 f_3-\partial_1 f_0
+\partial_2f_1-\partial_3 f_2)q^3\nonumber\\
\end{eqnarray}

\begin{eqnarray}
\partial_{z_3} F&=&
\frac{1}{4}(\partial_0 f_0+i\partial_1 f_1
-\partial_2f_2-i\partial_3 f_3)\nonumber\\
&+&\frac{1}{4}(\partial_0 f_1+i\partial_1 f_2
-\partial_2f_3-i\partial_3 f_0)q\nonumber\\
&+&\frac{1}{4}(\partial_0 f_2+i\partial_1 f_3
-\partial_2f_0-i\partial_3 f_1)q^2\nonumber\\
&+&\frac{1}{4}(\partial_0 f_3 +i\partial_1 f_0
-\partial_2f_1-i\partial_3 f_2)q^3\nonumber\\
\end{eqnarray}

In this case we can consider three types of holomoprphicity:
\begin{itemize}
\item{1. For  the first type of  holomorphicity  function $F( z_0
z_1,z_2, z_3 )$ we have the following three conditions:
\begin{eqnarray}
\frac{\partial F(z,z_1,z_2,z_3)}{\partial z_1} = \frac{\partial
F(z,z_1,z_2,z_3)}{\partial z_2 }=\frac{\partial
F(z,z_1,z_2,z_3)}{\partial z_3}=0.
\end{eqnarray}
}\\
\item{2. For  the second  type of  holomorphicity  function
$F(z,z_1,z_2,z_3)$ we can take  two conditions:
\begin{eqnarray}
\frac{\partial F(z,z_1,z_2,z_3)}{\partial z_2 z}=\frac{\partial
F(z,z_1,z_2,z_3)}{\partial z_3}=0.
\end{eqnarray}
}\\
\item{3. For  the third type of holomorphicity  function $F(
z,z_1,z_2,z_3)$ we can take just one condition:
\begin{eqnarray}
\frac{\partial F(z,z_1,z_2,z_3)}{\partial z }=0.
\end{eqnarray}
}\\
\end{itemize}

Similarly to the ternary case for $q^3=1$, one can get for $q^4=1$
 for the full Cauchi-Riemann system of the first type:

\begin{eqnarray}
\partial_0 f_0 &=&\partial_1 f_1=\partial_2 f_2= \partial_3 f_3\nonumber\\
\partial_3 f_0 &=&\partial_0 f_1=\partial_1 f_2= \partial_2 f_3\nonumber\\
\partial_2 f_0 &=&\partial_3 f_1=\partial_0 f_2= \partial_1 f_3 \nonumber\\
\partial_1 f_0 &=&\partial_2 f_1=\partial_3 f_2= \partial_0 f_3 \nonumber\\
\end{eqnarray}

and for quartic Laplace  equations one can easily get:

\begin{eqnarray}
&&\partial_0^4 f_p -\partial_1^4 f_p +\partial_2^4
f_p-\partial_3^4 f_p
 -2\partial_0^2\partial_2^2 f_p
 +2\partial_1^2\partial_3^2 f_p \nonumber\\
&& -4\partial_0^2\partial_1\partial_3f_p
+4\partial_1^2\partial_0\partial_2 f_p
-4\partial_2^2\partial_1\partial_3 f_p
+4\partial_3^2\partial_0\partial_2 f_p =0, \nonumber\\
\end{eqnarray}
where $p=0,1,2,3.$ These equations are invariant under
$T_4U(Abel)$ group symmetry, what it follows from $T_4U(Abel)$
invariance of the quartic form: $z z_1 z_2 z_3$.

Note, that the Caushi-Riemann conditions can be generalized for
any finite $C_n$ group for $q^n=1$:

We can consider the following  four   $4\times 4$  matrices from
$16$  $q$ matrices, which form the  quart-quaternion algebra (it
will be explained later):
\begin{eqnarray}
&&q_1= \left (
\begin{array}{cccc}
0 & 1 & 0 & 0\\
0 & 0 & 1 & 0\\
0 & 0 & 0 & 1\\
1 & 0 & 0 & 0\\
\end{array}
\right), \, q_2= \left (
\begin{array}{cccc}
0   & 1 & 0 & 0  \\
0   & 0 & j & 0  \\
0   & 0 & 0 & j^2 \\
j^3 & 0 & 0 & 0  \\
\end{array}
\right), \,q_3= \left (
\begin{array}{cccc}
0  & 1 & 0  & 0  \\
0  & 0 & j^2& 0  \\
0  & 0 & 0  & 1  \\
j^2& 0 & 0  & 0  \\
\end{array}
\right), q_4= \left (
\begin{array}{cccc}
0 & 1 & 0   & 0   \\
0 & 0 & j^3 & 0   \\
0 & 0 & 0   & j^2 \\
j & 0 & 0   & 0    \\
\end{array}
\right),
\nonumber\\
\end{eqnarray}
where $j=\exp{ 2 {\bf i} \pi/4}$.

These matrices satisfy to some  remarkable relations:
\begin{equation}
\{q_aq_bq_cq_d \}_{S_4}= \eta_{abcd} q_0
\end{equation}
with
\begin{eqnarray}
&&\eta_{1111}=-\eta_{2222}=\eta_{3333}=-\eta_{4444}=24 \nonumber\\
&&\eta_{1133}=-\eta_{2244}=2\nonumber\\
&&-\eta_{1123}=\eta_{1223}=-\eta_{2334}=\eta_{1344} =4\nonumber\\
\end{eqnarray}
where $j=\exp ( \pi/2)$ and $q_0$ is unit matrix. All others
tensor components $\eta_{...}$ are equal zero. Note that the
expression $\{q_aq_bq_cq_d \}_{S_4}$ contains the all possible
$24$ permutations of the $S_4$ symmetric group , for
$a=1,b=2,c=3,d=4$ ; $12$ for $a=b, c\neq d \neq a$ and etc.

Using these matrices one can get the quaternary Dirac equation:
\begin{eqnarray}
q_1 \frac{\partial \Psi}{\partial x_0}+ q_2 \frac{\partial
\Psi}{\partial x_1}+q_3 \frac{\partial \Psi}{\partial x_2} +q_4
\frac{\partial \Psi}{\partial x_3}=0,
\end{eqnarray}
where
\begin{eqnarray}
\Psi=(\psi_1,\psi_2,\psi_3,\psi_4),
\end{eqnarray}
is quartet of the wave functions, {\it i.e.} we introduced the
quaternary $1/4$ spin structure in $R^4$. The next
quaternary structures can appear in ${R}^{8,12,...}$
spaces.

We can consider the other set of  four   $4\times 4$  matrices
from $16$ $q$ matrices, which have the algebraic link to the first
set of the four matrices:

\begin{eqnarray}
&&q_9= \left (
\begin{array}{cccc}
0 & 0 & 0 & 1\\
1 & 0 & 0 & 0\\
0 & 1 & 0 & 0\\
0 & 0 & 1 & 0\\
\end{array}
\right), \, q_{10}= \left (
\begin{array}{cccc}
0     & 0     & 0   & 1  \\
j     & 0     & 0   & 0  \\
0     & j^2   & 0   & 0  \\
0     & 0     & j^3 & 0  \\
\end{array}
\right), \,q_{11}= \left (
\begin{array}{cccc}
0    & 0 & 0    & 1  \\
j^2  & 0 & 0    & 0  \\
0    & 1 & 0    & 0  \\
0    & 0 & j^2  & 0  \\
\end{array}
\right), q_{12}= \left (
\begin{array}{cccc}
0     & 0   & 0 & 1    \\
j^3   & 0   & 0 & 0   \\
0     & j^2 & 0 & 0    \\
0     & 0   & j & 0    \\
\end{array}
\right),
\nonumber\\
\end{eqnarray}

the second Dirac equation will be:

\begin{eqnarray}
q_9 \frac{\partial \Phi}{\partial x_0}+ q_{10} \frac{\partial
\Phi}{\partial x_1}+q_{11} \frac{\partial \Phi}{\partial x_2}
+q_{11} \frac{\partial \Phi}{\partial x_3}=0,
\end{eqnarray}
where
\begin{eqnarray}
\Phi=(\phi_1,\phi_2,\phi_3,\phi_4),
\end{eqnarray}
( here we are in process...)

 In order to diagonalize these equations we must act four
times with the same operator and we will get the above mentioned
quartic differential equation satisfied by each component
$\psi_l$, $l=1,2,3,4$.

The quartic  Laplace equations should be invariant under Abelian
three -parameter group $T_4U(Abel)$:

\begin{eqnarray}
z \rightarrow z^{\prime}= U z= \exp \{\phi_1 q + \phi_2 q^2
+\phi_3 q^3 \} z= U_1(\phi_1)U_2(\phi_2)U_3(\phi_3)
\end{eqnarray}

or in the coordinates $x_0,x_1,x_2,x_3$

\begin{eqnarray}
( x_0^{\prime}, x_1^{\prime},x_2^{\prime},x_3^{\prime})^t = O
\cdot (x_0,x_1,x_2,x_3^{\prime})^t,
\end{eqnarray}

where
\begin{eqnarray}
O_A=  \left (
\begin{array}{cccc}
    m_0  &     m_1  & m_2   & m_3   \\
    m_3  &     m_0  & m_1   & m_2  \\
    m_2  &     m_3  & m_0   & m_1      \\
    m_1  &     m_2  & m_3   & m_0      \\
\end{array}
\right ),
\end{eqnarray}
where
\begin{eqnarray}
Det O_A &=&m_0^4-m_1^4+m_2^4-m_3^4-2m_0^2m_2^2+2m_1^2m_3^2 \nonumber\\
 &&-4m_0^2m_1m_3+4m_1^2m_0m_2-4m_2^2m_1m_3+4m_3^2m_0m_2=1
 \nonumber\\
\end{eqnarray}

\section{${ C}_6$ complex numbers in  D=6}

Consider the ${C}_6$  complex numbers

\begin{eqnarray}
z=x_0q_0 +x_1 q+x_2q^2+x_3q^3+x_4q^4+x_5q^5
\end{eqnarray}
where we can consider two cases:

\begin{eqnarray}
A:\, q^6=q_0=1
\end{eqnarray}

and

\begin{eqnarray}
B: \,q^6=-q_0=-1.
\end{eqnarray}
Let define the conjugation operation of a new complex number:

\begin{eqnarray}
 \tilde q_0=q_0=1,  \qquad  \tilde q= j q, \qquad where \qquad j^{6}=1,
\end{eqnarray}
namely
\begin{eqnarray}
j =\exp{i \pi/3}.
\end{eqnarray}

Now one can  calculate the norm of this complex number:
\begin{eqnarray}
 z \tilde z \tilde{\tilde  z} \tilde {\tilde {\tilde z}}
 \tilde{\tilde {\tilde {\tilde z}}}
 \tilde{\tilde{\tilde {\tilde {\tilde z}}}}
 =1,
\end{eqnarray}
where

\begin{eqnarray}
z&=&x_0q_0 +x_1 q+x_2q^2+x_3q^3+x_4q^4+x_5q^5 \nonumber\\
\tilde z
&=&x_0q_0
 + x_1 \tilde  q
 + x_2 \tilde {q^2}
 + x_3 \tilde {q^3}
 + x_4 \tilde {q^4}
 + x_5 \tilde {q^5}
\nonumber\\
\tilde {\tilde z}
&=&x_0 \tilde {\tilde {q_0} }
 + x_1 \tilde {\tilde {q  } }
 + x_2 \tilde {\tilde {q^2} }
 + x_3 \tilde {\tilde {q^3} }
 + x_4 \tilde {\tilde {q^4} }
 + x_5 \tilde {\tilde {q^5} }
 \nonumber\\
\tilde {\tilde {\tilde z}}
&=&x_0 \tilde {\tilde  {\tilde {q_0}}}
 + x_1 \tilde {\tilde  {\tilde {q  }}}
 + x_2 \tilde {\tilde  {\tilde {q^2}}}
 + x_3 \tilde {\tilde  {\tilde {q^3}}}
 + x_4 \tilde {\tilde  {\tilde {q^4}}}
 + x_5 \tilde {\tilde  {\tilde {q^5}}}
 ,\nonumber\\
\tilde {\tilde {\tilde {\tilde {z}}}}
&=&x_0 \tilde {\tilde  {\tilde {\tilde {q_0}}}}
 + x_1 \tilde {\tilde  {\tilde {\tilde {q  }}}}
 + x_2 \tilde {\tilde  {\tilde {\tilde {q^2}}}}
 + x_3 \tilde {\tilde  {\tilde {\tilde {q^3}}}}
 + x_4 \tilde {\tilde  {\tilde {\tilde {q^4}}}}
 + x_5 \tilde {\tilde  {\tilde {\tilde {q^5}}}}
 ,\nonumber\\
 \tilde {\tilde {\tilde {\tilde {\tilde {z}}}}}
&=&x_0 \tilde {\tilde  {\tilde {\tilde {\tilde {q_0}}}}}
 + x_1 \tilde {\tilde  {\tilde {\tilde {\tilde {q  }}}}}
 + x_2 \tilde {\tilde  {\tilde {\tilde {\tilde {q^2}}}}}
 + x_3 \tilde {\tilde  {\tilde {\tilde {\tilde {q^3}}}}}
 + x_4 \tilde {\tilde  {\tilde {\tilde {\tilde {q^4}}}}}
 + x_5 \tilde {\tilde  {\tilde {\tilde {\tilde {q^5}}}}}
 ,\nonumber\\
 \end{eqnarray}
or
\begin{eqnarray}
z&=&x_0q_0 +     x_1  q +   x_2q^2 +    x_3q^3 +  x_4 q^4  +   x_5 q^5\nonumber\\
\tilde z
&=& x_0q_0 + j   x_1  q +j^2x_2q^2 +j^3 x_3q^3+j^4x_4q^4+j^5x_5q^5 \nonumber\\
\tilde {\tilde {z}}
&=& x_0q_0 + j^2 x_1  q +j^4x_2q^2 +j^0 x_3q^3+j^2x_4q^4+j^4x_5q^5 \nonumber\\
\tilde {\tilde {\tilde {z}}}
&=& x_0q_0 + j^3 x_1  q +j^0x_2q^2 +j^3 x_3q^3+j^0x_4q^4+j^3x_5q^5 \nonumber\\
\tilde {\tilde {\tilde {\tilde {z}}}}
&=& x_0q_0 + j^4 x_1  q +j^2x_2q^2 +j^0 x_3q^3+j^4x_4q^4+j^2x_5q^5  \nonumber\\
\tilde {\tilde {\tilde {\tilde {\tilde  {z}}}}}
&=&x_0q_0  + j^5 x_1  q +j^4x_2q^2 +j^3 x_3q^3+j^2x_4q^4+j  x_5q^5 \nonumber\\
\end{eqnarray}

We used the following relations:

\begin{eqnarray}
&&\tilde {q  } = j   q  ,                                        \qquad
  \tilde {q^2} = j^2 q^2,                                        \qquad
  \tilde {q^3} = j^3 q^3,                                        \qquad
  \tilde {q^4} = j^4 q^4,                                        \qquad
  \tilde {q^5} = j^5 q^5,                                        \nonumber\\
&&\tilde {\tilde {q  }}  = j^2 q  ,                              \qquad
  \tilde {\tilde {q^2}}  = j^4 q^2,                              \qquad
  \tilde {\tilde {q^3}}  = j^0 q^3,                              \qquad
  \tilde {\tilde {q^4}}  = j^2 q^4,                              \qquad
  \tilde {\tilde {q^5}}  = j^4 q^5,                              \nonumber\\
&&\tilde {\tilde {\tilde {q}}}  = j^3 q,                         \qquad
  \tilde {\tilde {\tilde {q^2}}}= j^0 q^2,                       \qquad
  \tilde {\tilde {\tilde {q^3}}}= j^3 q^3,                       \qquad
  \tilde {\tilde {\tilde {q^4}}}= j^0 q^4,                       \qquad
  \tilde {\tilde {\tilde {q^5}}}= j^3 q^5,                       \nonumber\\
&&\tilde {\tilde {\tilde {\tilde {q  }}}} = j^4 q,               \qquad
  \tilde {\tilde {\tilde {\tilde {q^2}}}} = j^2 q^2,             \qquad
  \tilde {\tilde {\tilde {\tilde {q^3}}}} = j^0 q^3,             \qquad\
  \tilde {\tilde {\tilde {\tilde {q^4}}}} = j^4 q^4,             \qquad\
  \tilde {\tilde {\tilde {\tilde {q^5}}}} = j^2 q^5,             \nonumber\\
&&\tilde {\tilde {\tilde {\tilde {\tilde {q  }}}}}  = j^5 q,     \qquad
  \tilde {\tilde {\tilde {\tilde {\tilde {q^2}}}}}  = j^4 q^2,   \qquad
  \tilde {\tilde {\tilde {\tilde {\tilde {q^3}}}}}  = j^3 q^3,   \qquad\
  \tilde {\tilde {\tilde {\tilde {\tilde {q^4}}}}}  = j^2 q^4,   \qquad\
  \tilde {\tilde {\tilde {\tilde {\tilde {q^5}}}}}  = j   q^5,   \nonumber\\
\end{eqnarray}

To find the equation of the surface we should take into account
the next identities:

\begin{eqnarray}
&&1+j+j^2+j^3+j^4+j^5=0   \nonumber\\
&&j+j^3+j^5=0,\, j-j^2=1,  \nonumber\\
&&1+j^2+j^4=0,\, j^5-j^4=1,\nonumber\\
\end{eqnarray}
or
\begin{eqnarray}
&&j  =\frac{1}{2} +i \frac{\sqrt{3}}{2},  \qquad
j^2=\frac{-1}{2}+i \frac{\sqrt{3}}{2}, \qquad
j^3=-1, \nonumber\\
&&j^4=\frac{-1}{2} -i \frac{\sqrt{3}}{2}, \qquad
j^5=\frac{1}{2}-i \frac{\sqrt{3}}{2}, \qquad
j^6=1.\nonumber\\
\end{eqnarray}

For the operations of conjugation one can use the other notations:

\begin{eqnarray}
\begin{array}{ccccccc}
z^{\{\tilde 0\}}=&x_0q_0 + &    x_1  q +  & x_2q^2   +  &  x_3q^3  +  &x_4 q^4   + &  x_5 q^5\\
z^{\{\tilde 1\}}
=& x_0q_0 + & j  x_1  q +  &j^2x_2q^2 +  &j^3 x_3q^3+  &j^4x_4q^4 + &j^5x_5q^5 \\
z^{\{\tilde 2\}}
=& x_0q_0 + &j^2 x_1  q +  &j^4x_2q^2 +  &j^0 x_3q^3+  &j^2x_4q^4 + &j^4x_5q^5 \\
z^{\{\tilde 3\}}
=& x_0q_0 + &j^3 x_1  q +  &j^0x_2q^2 +  &j^3 x_3q^3+  &j^0x_4q^4 + &j^3x_5q^5 \\
z^{\{\tilde 4\}}
=& x_0q_0 + &j^4 x_1  q +  &j^2x_2q^2 +  &j^0 x_3q^3+  &j^4x_4q^4 + &j^2x_5q^5  \\
z^{\{\tilde 5\}}
=&x_0q_0  + &j^5 x_1  q +  &j^4x_2q^2 +  &j^3 x_3q^3+  &j^2x_4q^4 + &j  x_5q^5 \\
\end{array}
\end{eqnarray}

\begin{eqnarray}
&&z[0]^6 - z[1]^6 + z[2]^6 - z[3]^6 + 6 z[2] z[3]^4 z[4] - \nonumber\\
&& 9 z[2]^2 z[3]^2 z[4]^2 + 2 z[2]^3 z[4]^3 + z[4]^6 - \nonumber\\
&& 6 z[2]^2 z[3]^3 z[5] + 12 z[2]^3 z[3] z[4] z[5] - \nonumber\\
&& 6 z[3] z[4]^4 z[5] - 3 z[2]^4 z[5]^2 + 9 z[3]^2 z[4]^2 z[5]^2 +\nonumber\\
&& 6 z[2] z[4]^3 z[5]^2 - 2 z[3]^3 z[5]^3 - 12 z[2] z[3] z[4] z[5]^3 + \nonumber\\
&& 3 z[2]^2 z[5]^4 - z[5]^6 - \nonumber\\
&& 3 z[0]^4 (z[3]^2 + 2 z[2] z[4] + 2 z[1] z[5]) + \nonumber\\
&& 3 z[1]^4 (z[4]^2 + 2 z[3] z[5]) + \nonumber\\
&& 3 z[1]^2 (3 z[2]^2 z[3]^2 + 2 z[2]^3 z[4] - z[4]^4 - \nonumber\\
&&    3 z[3]^2 z[5]^2 + 6 z[2] z[4] z[5]^2) - \nonumber\\
&& 2 z[1]^3 (z[3]^3 + 6 z[2] z[3] z[4] + 3 z[2]^2 z[5] + z[5]^3) + \nonumber\\
&& 2 z[0]^3 (z[2]^3 + z[4] (3 z[1]^2 + z[4]^2 + 6 z[3] z[5]) + \nonumber\\
&&    3 z[2] (2 z[1] z[3] + z[5]^2)) - \nonumber\\
&& 6 z[1] (z[2]^4 z[3] - 2 z[2] z[3] z[4]^3 + \nonumber\\
&&    3 z[2]^2 z[4]^2 z[5] + (z[4]^2 - z[3] z[5]) (z[3]^3 + z[5]^3)) -\nonumber\\
&& 3 z[0]^2 (2 z[1]^3 z[3] - z[3]^4 + 6 z[1] z[3] z[4]^2 +\nonumber\\
&&    3 z[1]^2 (z[2]^2 - z[5]^2) + 3 z[4]^2 (-z[2]^2 + z[5]^2) +\nonumber\\
&&    2 z[3] (3 z[2]^2 z[5] + z[5]^3)) + \nonumber\\
&& 6 z[0] (z[1]^4 z[2] + z[2]^3 z[3]^2 - z[2]^4 z[4] + \nonumber\\
&&    3 z[1]^2 z[3]^2 z[4] - 2 z[1]^3 z[4] z[5] - \nonumber\\
&&    z[2] (z[4]^4 - 3 z[3]^2 z[5]^2) + \nonumber\\
&&    z[4] (z[3]^2 z[4]^2 - 2 z[3]^3 z[5] + z[5]^4) + \nonumber\\
 &&   2 z[1] (z[2]^3 z[5] + z[4]^3 z[5] - z[2] (z[3]^3 + z[5]^3)))\nonumber\\
\end{eqnarray}
\begin{eqnarray}
&&x_0^6 - x_1^6 + 6 x_0 x_1^4 x_2 - 9 x_0^2 x_1^2 x_2^2 + 2 x_0^3 x_2^3 + x_2^6 - \nonumber\\
&& 6 x_0^2 x_1^3 x_3 + 12 x_0^3 x_1 x_2 x_3 - 6 x_1 x_2^4 x_3 - 3 x_0^4 x_3^2 +  \nonumber\\
&& 9 x_1^2 x_2^2 x_3^2 + 6 x_0 x_2^3 x_3^2 - 2 x_1^3 x_3^3 - 12 x_0 x_1 x_2 x_3^3 +  \nonumber\\
&& 3 x_0^2 x_3^4 - x_3^6 + 6 x_0^3 x_1^2 x_4 - 6 x_0^4 x_2 x_4 +  \nonumber\\
&& 6 x_1^2 x_2^3 x_4 - 6 x_0 x_2^4 x_4 - 12 x_1^3 x_2 x_3 x_4 +  \nonumber\\
&& 18 x_0 x_1^2 x_3^2 x_4 + 6 x_2 x_3^4 x_4 + 3 x_1^4 x_4^2 + 9 x_0^2 x_2^2 x_4^2 -  \nonumber\\
&& 18 x_0^2 x_1 x_3 x_4^2 - 9 x_2^2 x_3^2 x_4^2 - 6 x_1 x_3^3 x_4^2 +  \nonumber\\
&& 2 x_0^3 x_4^3 + 2 x_2^3 x_4^3 + 12 x_1 x_2 x_3 x_4^3 + 6 x_0 x_3^2 x_4^3 -  \nonumber\\
&& 3 x_1^2 x_4^4 - 6 x_0 x_2 x_4^4 + x_4^6 - 6 x_0^4 x_1 x_5 - 6 x_1^3 x_2^2 x_5 +  \nonumber\\
&& 12 x_0 x_1 x_2^3 x_5 + 6 x_1^4 x_3 x_5 - 18 x_0^2 x_2^2 x_3 x_5 - \nonumber\\
&& 6 x_2^2 x_3^3 x_5 + 6 x_1 x_3^4 x_5 - 12 x_0 x_1^3 x_4 x_5 +  \nonumber\\
&& 12 x_0^3 x_3 x_4 x_5 + 12 x_2^3 x_3 x_4 x_5 - 12 x_0 x_3^3 x_4 x_5 -  \nonumber\\
&& 18 x_1 x_2^2 x_4^2 x_5 + 12 x_0 x_1 x_4^3 x_5 - 6 x_3 x_4^4 x_5 +  \nonumber\\
&& 9 x_0^2 x_1^2 x_5^2 + 6 x_0^3 x_2 x_5^2 - 3 x_2^4 x_5^2 - 9 x_1^2 x_3^2 x_5^2 + \nonumber\\
&& 18 x_0 x_2 x_3^2 x_5^2 + 18 x_1^2 x_2 x_4 x_5^2 - 9 x_0^2 x_4^2 x_5^2 +  \nonumber\\
&& 9 x_3^2 x_4^2 x_5^2 + 6 x_2 x_4^3 x_5^2 - 2 x_1^3 x_5^3 - 12 x_0 x_1 x_2 x_5^3 -  \nonumber\\
&& 6 x_0^2 x_3 x_5^3 - 2 x_3^3 x_5^3 - 12 x_2 x_3 x_4 x_5^3 - 6 x_1 x_4^2 x_5^3 +  \nonumber\\
&& 3 x_2^2 x_5^4 + 6 x_1 x_3 x_5^4 + 6 x_0 x_4 x_5^4 - x_5^6 \nonumber\\
\end{eqnarray}

\begin{eqnarray}
&& ( x_0 - x_1 + x_2 - x_3 + x_4 - x_5),\nonumber\\
&&( x_0 + x_1 + x_2 + x_3 + x_4 + x_5) \nonumber\\
&& 1/2 (2 x_0 + x_1 - x_2 - 2 x_3 - x_4 + x_5 \nonumber\\
&&  - \sqrt{3} \sqrt{-x_1^2 - 2 x_1 x_2 - x_2^2 + 2 x_1 x_4 + 2 x_2 x_4
 -    x_4^2 + 2 x_1 x_5 + 2 x_2 x_5 - 2 x_4 x_5 - x_5^2)}), \nonumber\\
&&1/2 (2 x_0 + x_1 - x_2 - 2 x_3 - x_4 + x_5  \nonumber\\
&&+  \sqrt{3} \sqrt{(-x_1^2 - 2 x_1 x_2 - x_2^2 + 2 x_1 x_4 + 2 x_2 x_4 -
        x_4^2 + 2 x_1 x_5 + 2 x_2 x_5 - 2 x_4 x_5 - x_5^2)}), \nonumber\\
&& 1/2 (2 x_0 - x_1 - x_2 + 2 x_3 - x_4 - x_5 \nonumber\\
&&-    \sqrt{3} \sqrt{(-x_1^2 + 2 x_1 x_2 - x_2^2 - 2 x_1 x_4 + 2 x_2 x_4 -
       x_4^2 + 2 x_1 x_5 - 2 x_2 x_5 + 2 x_4 x_5 - x_5^2)}),               \nonumber\\
&& 1/2 (2 x_0 - x_1 - x_2 + 2 x_3 - x_4 - x_5                  \nonumber\\
&&+   \sqrt{3} \sqrt{(-x_1^2 + 2 x_1 x_2 - x_2^2 - 2 x_1 x_4 + 2 x_2 x_4 -
       x_4^2 + 2 x_1 x_5 - 2 x_2 x_5 + 2 x_4 x_5 - x_5^2)})          \nonumber\\
\end{eqnarray}
In the case $A$ the ${C}_6$ the unit complex numbers define the surface
 which can be factorized:

\begin{eqnarray}
&&(x_0+x_1+x_2+x_3+x_4+x_5+x_6)(x_0-x_1+x_2-x_3+x_4-x_5+x_6) \nonumber\\
&&[(x_0^2+x_1^2+x_2^2+x_3^2+x_4^2+x_5^2
 -x_0x_2-x_0x_4-x_1x_3-x_1x_5-x_2x_4-x_3x_5)   \nonumber\\
&+&(x_0x_1-2x_0x_3+x_0x_5+x_1x_2-2x_1x_4+x_2x_3-2x_2x_5+x_3x_4+x_4x_5)]
\nonumber\\
&&[(x_0^2+x_1^2+x_2^2+x_3^2+x_4^2+x_5^2
 -x_0x_2-x_0x_4-x_1x_3-x_1x_5-x_2x_4-x_3x_5)   \nonumber\\
&-&(x_0x_1-2x_0x_3+x_0x_5+x_1x_2-2x_1x_4+x_2x_3-2x_2x_5+x_3x_4+x_4x_5)]
\nonumber\\
\end{eqnarray}

This surface is invariant under the following transformations:

\begin{eqnarray}
(x_0^{\prime}, x_1^{\prime},x_2^{\prime},x_3^{\prime},x_4^{\prime},x_5^{\prime})=O(A) (x_0,x_1,x_2,x_3,x_4,x_5),
\end{eqnarray}
where

\begin{eqnarray}
O(A)=
\left (
\begin{array}{cccccc}
m_0&m_1&m_2&m_3&m_4&m_5 \\
m_5&m_0&m_1&m_2&m_3&m_4 \\
m_4&m_5&m_0&m_1&m_2&m_3 \\
m_3&m_4&m_5&m_0&m_1&m_2 \\
m_2&m_3&m_4&m_5&m_0&m_1 \\
m_1&m_2&m_3&m_4&m_5&m_0 \\
\end{array}
\right )
\end{eqnarray}
where $Det O(A)=1$. The expressions for the multi-sin functions one
can get through the ${ C}_6$ Euler formul:

\begin{equation}
\exp {( \phi_1 q + \phi_2 q^2 + \phi_3 q^3 +\phi_4 q^4 \phi_5 q^5)}=
m_0( \phi_1,...,\phi_5) q + ...+m_5(\phi_0,...,\phi_5) q^5.
\end{equation}

In the case $B$ the ${ C}_6$ unit complex numbers define the following surface:
\begin{eqnarray}
\left(
\begin{array}{cccccc}
x_0 & -x_1 & -x_2 & -x_3 & -x_4 & -x_5 \\
x_5 &  x_0 & -x_1 & -x_2 & -x_3 & -x_4 \\
x_4 &  x_5 &  x_0 & -x_1 & -x_2 & -x_3\\
x_3 &  x_4 &  x_5 &  x_0 & -x_1 & -x_2\\
x_2 &  x_3 &  x_4 &  x_5 &  x_0 & -x_1\\
x_1 &  x_2 &  x_3 &  x_4 &  x_5 &  x_0\\
\end{array}
\right)
\end{eqnarray}

\begin{eqnarray}
&&x_0^6 + x_1^6 + 6 x_0 x_1^4 x_2 + 9 x_0^2 x_1^2 x_2^2 + 2 x_0^3 x_2^3 + x_2^6 + \nonumber\\
&& 6 x_0^2 x_1^3 x_3 + 12 x_0^3 x_1 x_2 x_3 - 6 x_1 x_2^4 x_3 + 3 x_0^4 x_3^2 + \nonumber\\
&& 9 x_1^2 x_2^2 x_3^2 - 6 x_0 x_2^3 x_3^2 - 2 x_1^3 x_3^3 + 12 x_0 x_1 x_2 x_3^3 + \nonumber\\
&& 3 x_0^2 x_3^4 + x_3^6 + 6 x_0^3 x_1^2 x_4 + 6 x_0^4 x_2 x_4 + \nonumber\\
&& 6 x_1^2 x_2^3 x_4 + 6 x_0 x_2^4 x_4 - 12 x_1^3 x_2 x_3 x_4 - \nonumber\\
&& 18 x_0 x_1^2 x_3^2 x_4 - 6 x_2 x_3^4 x_4 + 3 x_1^4 x_4^2 + 9 x_0^2 x_2^2 x_4^2 - \nonumber\\
&& 18 x_0^2 x_1 x_3 x_4^2 + 9 x_2^2 x_3^2 x_4^2 + 6 x_1 x_3^3 x_4^2 - \nonumber\\
&& 2 x_0^3 x_4^3 - 2 x_2^3 x_4^3 - 12 x_1 x_2 x_3 x_4^3 + 6 x_0 x_3^2 x_4^3 + \nonumber\\
&& 3 x_1^2 x_4^4 - 6 x_0 x_2 x_4^4 + x_4^6 + 6 x_0^4 x_1 x_5 - 6 x_1^3 x_2^2 x_5 - \nonumber\\
&& 12 x_0 x_1 x_2^3 x_5 + 6 x_1^4 x_3 x_5 - 18 x_0^2 x_2^2 x_3 x_5 + \nonumber\\
&& 6 x_2^2 x_3^3 x_5 - 6 x_1 x_3^4 x_5 + 12 x_0 x_1^3 x_4 x_5 - \nonumber\\
&& 12 x_0^3 x_3 x_4 x_5 - 12 x_2^3 x_3 x_4 x_5 - 12 x_0 x_3^3 x_4 x_5 + \nonumber\\
&& 18 x_1 x_2^2 x_4^2 x_5 + 12 x_0 x_1 x_4^3 x_5 - 6 x_3 x_4^4 x_5 + \nonumber\\
&& 9 x_0^2 x_1^2 x_5^2 - 6 x_0^3 x_2 x_5^2 + 3 x_2^4 x_5^2 + 9 x_1^2 x_3^2 x_5^2 + \nonumber\\
&& 18 x_0 x_2 x_3^2 x_5^2 - 18 x_1^2 x_2 x_4 x_5^2 + 9 x_0^2 x_4^2 x_5^2 +\nonumber\\
&& 9 x_3^2 x_4^2 x_5^2 + 6 x_2 x_4^3 x_5^2 + 2 x_1^3 x_5^3 - 12 x_0 x_1 x_2 x_5^3 + \nonumber\\
&& 6 x_0^2 x_3 x_5^3 - 2 x_3^3 x_5^3 - 12 x_2 x_3 x_4 x_5^3 - 6 x_1 x_4^2 x_5^3 + \nonumber\\
&& 3 x_2^2 x_5^4 + 6 x_1 x_3 x_5^4 - 6 x_0 x_4 x_5^4 + x_5^6\nonumber\\
\end{eqnarray}

\begin{eqnarray}
&&z[0]^6 + z[1]^6 + z[2]^6 + z[3]^6 - 6 z[2] z[3]^4 z[4] + \nonumber\\
&& 9 z[2]^2 z[3]^2 z[4]^2 - 2 z[2]^3 z[4]^3 + z[4]^6 + \nonumber\\
&& 6 z[2]^2 z[3]^3 z[5] - 12 z[2]^3 z[3] z[4] z[5] - \nonumber\\
&& 6 z[3] z[4]^4 z[5] + 3 z[2]^4 z[5]^2 + 9 z[3]^2 z[4]^2 z[5]^2 + \nonumber\\
&& 6 z[2] z[4]^3 z[5]^2 - 2 z[3]^3 z[5]^3 - 12 z[2] z[3] z[4] z[5]^3 +\nonumber\\
&& 3 z[2]^2 z[5]^4 + z[5]^6 + \nonumber\\
&& 3 z[0]^4 (z[3]^2 + 2 z[2] z[4] + 2 z[1] z[5]) + \nonumber\\
&& 3 z[1]^4 (z[4]^2 + 2 z[3] z[5]) + \nonumber\\
&& 3 z[1]^2 (3 z[2]^2 z[3]^2 + 2 z[2]^3 z[4] + z[4]^4 + \nonumber\\
&&    3 z[3]^2 z[5]^2 - 6 z[2] z[4] z[5]^2) -\nonumber\\
&& 2 z[1]^3 (z[3]^3 + 6 z[2] z[3] z[4] + 3 z[2]^2 z[5] - z[5]^3) -\nonumber\\
&& 2 z[0]^3 (z[2]^3 - z[4] (-3 z[1]^2 + z[4]^2 + 6 z[3] z[5]) + \nonumber\\
&&    z[2] (6 z[1] z[3] - 3 z[5]^2)) - \nonumber\\
&& 6 z[1] (z[2]^4 z[3] + 2 z[2] z[3] z[4]^3 - \nonumber\\
&&    3 z[2]^2 z[4]^2 z[5] + (-z[4]^2 + z[3] z[5]) (z[3]^3 - z[5]^3)) + \nonumber\\
&& 3 z[0]^2 (2 z[1]^3 z[3] + z[3]^4 - 6 z[1] z[3] z[4]^2 + \nonumber\\
&&    3 z[1]^2 (z[2]^2 + z[5]^2) + 3 z[4]^2 (z[2]^2 + z[5]^2) + \nonumber\\
&&    z[3] (-6 z[2]^2 z[5] + 2 z[5]^3)) - \nonumber\\
&& 6 z[0] (z[1]^4 z[2] - z[2]^3 z[3]^2 + z[2]^4 z[4] - \nonumber\\
&&    3 z[1]^2 z[3]^2 z[4] + 2 z[1]^3 z[4] z[5] - \nonumber\\
&&    z[2] (z[4]^4 - 3 z[3]^2 z[5]^2) + \nonumber\\
&&    z[4] (z[3]^2 z[4]^2 - 2 z[3]^3 z[5] - z[5]^4) - \nonumber\\
&&    2 z[1] (z[2]^3 z[5] - z[4]^3 z[5] + z[2] (-z[3]^3 + z[5]^3)))\nonumber\\
\end{eqnarray}

\begin{eqnarray}
z z^{\{\tilde 1\}}z^{\{\tilde 2\}}z^{\{\tilde 3\}}z^{\{\tilde 4\}}z^{\{\tilde 5\}}
&=&x_0^6+x_1^6+x_2^6+x_3^6+x_4^6+x_5^6                                    \nonumber\\
&+&x_0^4x_3^2+x_1^4x_4^2+3x_2^4x_5^2                                       \nonumber\\
&+&x_3^4x_0^2+ 3x_4^4x_1^2+3 x_5^4x_2^2                                       \nonumber\\
&+&x_0^4(x_2x_4+x_1x_5)
 + x_1^4(x_0x_2+x_3x_5)                                                    \nonumber\\
&+& x_2^4(x_0x_4+x_1x_3)
 + x_3^4(x_1x_5+x_2x_4)                                                    \nonumber\\
&+& x_4^4(x_0x_2+x_3x_5)
 + x_5^4(x_0x_4+x_1x_3)                                                   \nonumber\\
&+&x_0^3x_2^3+x_2^3x_4^3+x_4^3x_0^3+x_1^3x_3^3+x_3^3x_5^3+x_5^3x_1^3      \nonumber\\
&+&x_0^3(x_1^2x_4+x_5^2x_2+x_1x_2x_3+x_3x_4x_5)                           \nonumber\\
&+& x_1^3(x_0^2x_3+x_2^2x_5+x_2x_3x_4+x_0x_4x_5)                          \nonumber\\
&+&x_2^3(x_1^2x_4+x_3^2x_0+x_0x_1x_5+x_3x_4x_5)                            \nonumber\\               &+&x_3^3(x_2^2x_5+x_4^2x_1+x_0x_1x_2+x_0x_4x_5)                            \nonumber\\
&+&x_4^3(x_3^2x_0+x_5^2x_2+x_1x_2x_3+x_0x_1x_5)                             \nonumber\\
&+&x_5^3(x_4^2x_1+x_0^2x_3+x_0x_1x_2+x_2x_3x_4)                            \nonumber\\
&+&x_0^2x_1^2x_2^2
 + x_0^2x_2^2x_4^2
 + x_0^2x_1^2x_5^2
 + x_0^2x_4^2x_5^2                                                         \nonumber\\
 &+& x_1^2x_2^2x_3^2
 +x_1^2x_3^2x_5^2
 + x_2^2x_3^2x_4^2
 + x_3^2x_4^2x_5^2                                                        \nonumber\\
&+&x_0^2x_2^2x_3x_5
 + x_0^2x_3^2x_1x_5
 + x_0^2x_3^2x_2x_4
 + x_0^2x_4^2x_1x_3
 + x_1^2x_3^2x_0x_4\nonumber\\
 &+& x_1^2x_5^2x_2x_4
 + x_1^2x_4^2x_3x_5
 + x_2^2x_4^2x_1x_5
 + x_2^2x_5^2x_0x_4
 + x_3^2x_5^2x_0x_2                                   \nonumber\\
&+&x_0^2x_1x_2x_4x_5
 + x_1^2x_0x_2x_3x_5
 + x_2^2x_0x_1x_3x_4\nonumber\\
 &+& x_3^2x_1x_2x_4x_5
 + x_4^2x_0x_2x_3x_5
 + x_5^2x_0x_1x_3x_4 =1                               \nonumber\\
\end{eqnarray}

This surface is invariant under
transformations:
\begin{eqnarray}
(x_0^{\prime}, x_1^{\prime},x_2^{\prime},x_3^{\prime},x_4^{\prime},x_5^{\prime})=O(B) (x_0,x_1,x_2,x_3,x_4,x_5),
\end{eqnarray}
where

\begin{eqnarray}
O(B)=
\left (
\begin{array}{cccccc}
m_0&-m_1&m_2&-m_3&m_4&-m_5 \\
m_5&m_0&-m_1&m_2&-m_3&m_4 \\
-m_4&m_5&m_0&-m_1&m_2&-m_3 \\
m_3&-m_4&m_5&m_0&-m_1&m_2 \\
-m_2&m_3&-m_4&m_5&m_0&-m_1 \\
m_1&-m_2&m_3&-m_4&m_5&m_0 \\
\end{array}
\right )
\end{eqnarray}
where $Det O(B)=1$.

\section{|Conclusions}
Thus using $C_n$ group and such matrices  $(A and B)$ there is a very natural way to extend the complexification for all $R^n $ Euclidean spaces:
to get   and to study the properties of analicity function, n-dimensional Laplace euation, to introduce new Dirac equation with spin $1/n$,
n-dimensional Pithagore theorem, to find (n-1)-parameter extension of Abelian $U(1)$ group. The last could be related with multi extension theory of Light.
At last we gave
 solutions of the Phiafagore  Equations for 3D case in integer numbers.
(This table has been calculated by Samoilenko.
\begin{eqnarray}
\begin{array}{|c|c|c|c|c|} \hline
  a &  b & c &   d  &    d^3    \\ \hline
   2 &  3 & 3 &   2  &    8      \\
  2 &  3 & 4 &   3  &   27      \\
  3 & 19 & 27&   28 &  21952    \\
  3 & 31 & 38&   42 &  74088    \\
  4 &  6 &  6&    4 &   64  \\
  4 &  6 &  8&    6 &  216  \\
  5 & 25 & 42&    42&   74088  \\
  6 &  9 &  9&    6 &   216  \\
  6 &  9 & 12&    9 &  729 \\
\end{array}
\end{eqnarray}
--------------------------------------------------------------

Acknowledments.
We are very grateful   to Igor Ajinenko, Luis Alvarez-Gaume, Ignatios Antoniadis, Genevieve Belanger, Nikolai Boudanov,
 Tatjana Faberge, M. Vittoria Garzelli,
 Nanie Perrin, Alexander Poukhov, for very nice  support.
Some important results we  have got from very  nice discussions with
Alexey Dubrovskiy, John Ellis, Lev Lipatov, Richard Kerner,
Andrey Koulikov, Michel Rausch de Traunberg,
Robert Yamaleev.
Thank them very much.

\end{document}